\documentclass[12pt,english,preprint]{revtex4-1}
\usepackage{times}
\usepackage{array}
\usepackage{longtable}
\usepackage{float}
\usepackage{amsmath}
\usepackage{bm}
\usepackage{graphicx}
\usepackage{amssymb}
\usepackage{color}
\usepackage[FIGTOPCAP]{subfigure}

\makeatletter
\usepackage{fancyhdr}
\usepackage{babel}
\makeatother
\begin{document}

\title{An Efficient Surrogate Model of Secondary Electron Formation and Evolution}

\author{Christopher J. McDevitt}
\email{cmcdevitt@ufl.edu}
\affiliation{Nuclear Engineering Program, Department of Materials Science and Engineering, University of Florida, Gainesville, FL 32611, United States of America}
\author{Jonathan Arnaud}
\affiliation{Nuclear Engineering Program, Department of Materials Science and Engineering, University of Florida, Gainesville, FL 32611, United States of America}
\author{Xian-Zhu Tang}
\affiliation{Theoretical Division, Los Alamos National Laboratory, Los Alamos, NM 87545, United States of America}

\date{\today}

\begin{abstract}

This work extends the adjoint-deep learning framework for runaway
electron (RE) evolution developed in Ref. \cite{mcdevittpart12024} to
account for large-angle collisions. By incorporating large-angle
collisions the framework allows the avalanche of REs to be captured, an
essential component to RE dynamics. This extension is accomplished by
using a Rosenbluth-Putvinski approximation to estimate the
distribution of secondary electrons generated by large-angle
collisions. By evolving both the primary and multiple generations of secondary
electrons, the present formulation is able to capture both
the detailed temporal evolution of a RE population beginning from an
arbitrary initial momentum space distribution, along with providing
approximations to the saturated growth and decay rates of the RE
population. Predictions of the adjoint-deep learning framework are
verified against a traditional RE solver, with good agreement present
across a broad range of parameters.

\end{abstract}

\maketitle


\section{Introduction}

The description of runaway electron (RE) formation and evolution in magnetized fusion devices poses a substantial scientific challenge. While reduced analytic~\cite{Kruskal-Bernstein:1962, Connor:1975, Rosenbluth:1997, smith2008hot, Martin:2017}, semi-analytic~\cite{Aleynikov:2015, hesslow2019influence, mcdevitt2018relation} or machine learning~\cite{hesslow2019evaluation, McDevitt:hottail:2023, yang2024pseudoreversible, arnaud2024physics} models are often used to infer RE generation rates, threshold electric fields, or the number of seed REs, such models are unable to recover the detailed dynamics of RE evolution. Instead, the description of RE dynamics is often performed using continuum~\cite{Harvey:2000, Nilsson:2015, guo-etal-pop-2019,hoppe2021dream,Rudi-etal-JCP-2024} or particle-based RE solvers~\cite{Eriksson:2003, Sommariva:2017, mcdevitt2019avalanche, beidler2024wall}.
In a companion paper~\cite{mcdevittpart12024}, the decay rate of a population of primary electrons was evaluated using a parametric solution to the adjoint of the relativistic Fokker-Planck equation. By targeting the adjoint to the relativistic Fokker-Planck equation, a single solution to this equation for a given set of parameters allowed the RE density to be evolved forward in time beginning from an arbitrary initial momentum space distribution of electrons. The framework was further generalized by utilizing a physics-informed neural network (PINN) to identify a \emph{parametric} solution to the adjoint of the relativistic Fokker-Planck equation. Hence, once trained the combined adjoint-deep learning framework enabled the dynamics of the RE density to be inferred beginning from an arbitrary initial momentum space distribution for a range of physics parameters. While the offline training time of the PINN was substantially greater than the solution time of a traditional RE solver, its online inference time was orders of magnitude faster, suggesting its potential as a surrogate model for RE evolution in integrated descriptions of tokamak disruptions.


Reference \cite{mcdevittpart12024}, however, did not describe the exponential growth of the RE population due to the avalanche mechanism~\cite{Sokolov:1979}, an essential component when describing the conversion of Ohmic to RE current in tokamak devices~\cite{Hender:2007, breizman2019physics}. The present paper seeks to remove this limitation by extending the adjoint-deep learning framework to incorporate large-angle collisions, thus allowing for the amplification of an initial seed population. A second generalization of the framework will be achieved by training across a broader range of physics parameters compared to Ref. \cite{mcdevittpart12024}. This is accomplished by introducing a normalized time and energy coordinate that allows features of the runaway probability function (RPF) to be accurately tracked while using a comparable number of training points as the narrow parameter range employed in Ref. \cite{mcdevittpart12024}.
The derived adjoint-deep learning approach will thus provide an efficient framework through which RE dynamics can be accurately described.

The present paper is not the first to evaluate the avalanche growth
rate of REs using an adjoint approach~\cite{Liu:2017}, nor the first
to use a PINN for evaluating the avalanche growth
rate~\cite{arnaud2024physics}. With regard to
Ref. \cite{arnaud2024physics}, that paper employed a PINN to solve the
steady state adjoint of the relativistic Fokker-Planck equation. Using
that solution, the saturated avalanche growth rate of the RE
population was inferred. A limitation of Ref. \cite{arnaud2024physics},
however, was that it computed the rate that secondary electrons are
generated and run away, without accounting for the decay of the
primary distribution. Thus, when below threshold
Ref. \cite{arnaud2024physics} simply predicted a negligibly small
avalanche growth, without accounting for the decay of the overall RE
population. This resulted in a slight ambiguity in the determination of the RE avalanche threshold, together with the inability to correctly describe RE decay when below threshold. Furthermore, Ref. \cite{arnaud2024physics} focused on the saturated growth rate of REs, and thus did not treat the transient evolution of the RE population. Results from the present model will fill these gaps by providing an efficient
means of evaluating the temporal evolution of the RE density both above and below
threshold. The near and below threshold limits are of particular
importance during the RE plateau, where accounting for the decay of
the RE distribution is essential for accurately predicting the
dissipation of the RE beam. While analytic approximations to this rate
are available, see for example Ref. \cite{Rosenbluth:1997}, such
formulae are
highly inaccurate when below threshold~\cite{mcdevittpart12024}. When carrying out these extensions we will consider the limit of a completely screened plasma for a constant electric field. A more comprehensive treatment of REs during a disruption will require including a temporally varying electric field and partial screening corrections~\cite{Hesslow:2017}, which will be left to a future work. 

The remainder of this paper is organized as follows. Section \ref{sec:AFR} discusses an extension of the adjoint-deep learning formulation of the relativistic Fokker-Planck equation to include a secondary source term. Physics-informed neural networks are briefly described in Sec. \ref{sec:PINN}. The temporal evolution of the number density of an avalanching RE population both above and below threshold is given in Sec. \ref{sec:RDE}. Section \ref{sec:PVR} evaluates the avalanche growth and decay rates across a broad range of plasma conditions and compares with a traditional RE solver. A brief discussion along with conclusions are given in Sec. \ref{sec:C}.

\section{\label{sec:AFR}Adjoint Framework for Relativistic Electron Evolution}

\subsection{\label{sec:}The Adjoint of the Relativistic Fokker-Planck Equation}

This section will provide a brief description of an adjoint formulation for evolving the RE density beginning from an arbitrary initial momentum space distribution. A more detailed description is given in Ref. \cite{mcdevittpart12024}. The adjoint of the relativistic Fokker Planck equation can be written as~\cite{karney1986current, Liu:2017, zhang2017backward, McDevitt:hottail:2023}:
\begin{subequations}
\label{eq:TDRP9}
\begin{align}
\frac{\partial P}{\partial t} - E^* \left( P, t\right) - C^* \left( P, t\right) - R^* \left( P, t\right) = 0
, \label{eq:TDRP9a}
\end{align}
with the adjoint operators defined by:
\begin{equation}
E^* \left( P, t\right) = E_\Vert \xi \frac{\partial P}{\partial p} + \left( \frac{1-\xi^2}{p}\right) E_\Vert \frac{\partial P}{\partial \xi}
, \label{eq:TDRP9b}
\end{equation}
\begin{equation}
C^* \left( P, t\right) = C_F \frac{\partial P}{\partial p} - \frac{C_B}{p^2} \frac{\partial}{\partial \xi} \left[ \left( 1-\xi^2\right) \frac{\partial P}{\partial \xi} \right]
, \label{eq:TDRP9c}
\end{equation}
\begin{equation}
R^* \left( P, t\right) = \alpha p \gamma \left( 1 - \xi^2 \right) \frac{\partial P}{\partial p} - \alpha \frac{\xi \left( 1-\xi^2 \right)}{\gamma} \frac{\partial P}{\partial \xi}
. \label{eq:TDRP9d}
\end{equation}
\end{subequations}
Here, we have defined the electron's pitch by $\xi \equiv p_\Vert / p$ $\in$ $\left[ -1, 1 \right]$, time is normalized as $t \to t/\tau_c$, with $\tau_c \equiv 4\pi \epsilon^2_0 m^2_e c^3 / \left( e^4 n_e \ln \Lambda \right)$ and Coulomb logarithm $\ln\Lambda$, momentum as $p\to p / \left( m_e c \right)$, the electric field as $E_\Vert \to E_\Vert / E_c$, where $E_c \equiv m_e c / \left( e\tau_c\right)$ is the Connor-Hastie electric field~\cite{Connor:1975}, the Lorentz factor is defined as $\gamma = \sqrt{1+p^2}$, $\alpha \equiv \tau_c/\tau_s$, and $\tau_s \equiv 6\pi \epsilon_0 m^3_e c^3 / \left( e^4 B^2\right)$ is the timescale associated with synchrotron radiation. The coefficients for the collisional drag $C_F$ and pitch-angle scattering $C_B$ in Eq. (\ref{eq:TDRP9c}) are defined by:
\begin{subequations}
\begin{equation}
C_F \equiv \frac{\gamma^2}{p^2}
, \label{eq:TDRP5a}
\end{equation}
\begin{equation}
C_B \equiv \frac{\gamma}{2} \frac{  \left( Z_{eff} + 1 \right)}{p}
. \label{eq:TDRP5b}
\end{equation}
\end{subequations}
An adjoint problem for the runaway probability function (RPF) $P$ can be defined by enforcing the terminal condition and momentum boundary conditions:
\begin{subequations}
\label{eq:RPF1}
\begin{equation}
P \left( t=t_{final} \right) = \Theta \left( p - p_{RE}\right)
, \label{eq:RPF1a}
\end{equation}
\begin{equation}
P \left( p=p_{min} \right) = 0
, \label{eq:RPF1b}
\end{equation}
\begin{equation}
P \left( p=p_{max} \right) =
\begin{cases}
1, & U_p \left( p=p_{max} \right) > 0 \\
\text{unconstrained}, & U_p \left( p=p_{max} \right) < 0 
\end{cases}
, \label{eq:RPF1c}
\end{equation}
\end{subequations}
where $\Theta \left( x\right)$ is a Heaviside function and $U_p$ is defined by:
\begin{equation}
U_p \equiv -E_\Vert \xi - C_F - \alpha p \gamma \left( 1 - \xi^2 \right)
. \label{eq:TDRP8}
\end{equation}
Here, $p_{RE}$ indicates the momentum above which an electron will be counted as a RE, whose specific value will be discussed below. Once the adjoint problem defined by Eq. (\ref{eq:TDRP9}) with the terminal and boundary conditions defined by Eq. (\ref{eq:RPF1}) has been solved, the RE density at a time between $0 \leq t \leq t_{final}$ is given by~\cite{mcdevittpart12024}:
\begin{equation}
n_{RE} \left( t \right) = \int d^3 p f^{(init)}_e \left( p, \xi \right) P \left( p, \xi, \tau ; t_{final} \right)
, \label{eq:TDRP20}
\end{equation}
where $\tau \equiv t_{final} - t$. This latter formulation will allow the time evolution of the density moment of an arbitrary initial momentum space distribution of REs $f^{(init)}_e \left( p, \xi \right)$ to be projected to any time between $0$ and $t_{final}$. It does not, however, include large-angle collisions which are essential when describing RE populations for the large electric fields expected in tokamak disruptions. The extension to include large-angle collisions is discussed in Sec. \ref{sec:REA} below.

\subsection{\label{sec:REA}Runaway Electron Avalanche}

Once the time evolution of the RE density is inferred, this can be used to evaluate the rate that a given seed population of REs will be enhanced by the avalanche mechanism. From Eq. (\ref{eq:TDRP20}), the evolution of the RE seed is given by:
\begin{equation}
n_{seed} \left( t \right) = \int d^3 p f^{(seed)}_e \left( p, \xi \right) P \left( p, \xi, \tau ; t_{final} \right)
. \label{eq:REA1}
\end{equation}
This seed population will generate additional REs via large-angle collisions. By introducing a temporal grid from $t=0$ to $t=t_{final}$, in increments of $\Delta t_{av} \equiv t_{final} / N_{av}$, where $N_{av}$ is the number of avalanche time steps, 
the distribution of secondary electrons after one time step can be approximated by: 
\begin{equation}
f^{(1)}_{sec} \left( p, \xi, t_1 \right) = \Delta t_{av} S \left( p, \xi, t_1 \right)
, \label{eq:REA2}
\end{equation}
where $t_i = i \Delta t_{av}$, $f^{(1)}_{sec}$ indicates the first generation of secondary electrons, and $S\left( p, \xi \right)$ is the source of secondary electrons defined by:
\begin{equation}
S \left( p, \xi, t_1 \right) = \int d^3 p^\prime S_0 \left( p^\prime, \xi^\prime, p, \xi \right) f_e \left( p^\prime, \xi^\prime, t_1 \right)
. \label{eq:REA2sub1}
\end{equation}
Here $f_e \left( p^\prime, \xi^\prime, t_{1} \right)$ is the seed electron population evaluated after one time step, and $S_0 \left( p^\prime, \xi^\prime, p, \xi \right)$ is given by:
\begin{equation}
S_0 \left( p^\prime, \xi^\prime, p, \xi \right) = n_e cr^2_e \frac{v^\prime}{2\pi p^2} \frac{d \sigma_M \left( p^\prime, p \right)}{d p} \Pi \left( p^\prime, \xi^\prime, p; \xi \right)
, \label{eq:REA3}
\end{equation}
where $r_e = e^2/\left( 4\pi\epsilon_0 m_e c^2\right)$ is the classical electron radius, $d \sigma_M/dp$ is the M\o ller cross section~\cite{Moller:1932,Ashkin:1954}, $\Pi \left( p^\prime, \xi^\prime , p ; \xi \right)$ describes the pitch-angle dependence of secondary electron generation (see Ref. \cite{Boozer:2015} for an explicit expression), $d^3p = 2\pi p^2 dp d\xi$, all variables have been dedimensionalized according to $p^\prime \to p^\prime/m_ec$, $v^\prime \to v^\prime/c$, and $\sigma_M \to \sigma_M / r^2_e$. Since the adjoint formulation does not immediately give us the time evolution of the seed electron distribution $f_e \left( p^\prime, \xi^\prime, t \right)$, only its density moment, we will need to introduce a closure. The simplest closure available, introduced in Ref. \cite{Rosenbluth:1997}, takes the limit whereby the energy of the primary electrons is assumed to be infinite $p^\prime \to \infty$ and the pitch is taken to be aligned with the magnetic field $\xi^\prime = -1$. Taking these limits, Eq. (\ref{eq:REA2sub1}) reduces to:
\begin{equation}
S \left( p, \xi, t_1 \right)  = n_e n_{RE} \left( t_1 \right) c r^2_e \frac{v}{\gamma^2-1} \frac{1}{\left( \gamma-1 \right)^2} \delta \left( \xi - \xi_1 \right)
, \label{eq:REA4}
\end{equation}
where $\xi_1$ is defined by:
\begin{equation}
\xi_1 = -\sqrt{\frac{\gamma-1}{\gamma+1}}
. \label{eq:REA5}
\end{equation}
Substituting Eq. (\ref{eq:REA4}) into (\ref{eq:REA2}) yields an explicit expression for the momentum space distribution of the first generation of secondaries, i.e. 
\begin{equation}
f^{(1)}_{sec} \left( p, \xi \right) = \Delta t_{av} n_e n_{RE} \left( t_1 \right) c r^2_e \frac{v}{\gamma^2-1} \frac{1}{\left( \gamma-1 \right)^2} \delta \left( \xi - \xi_1 \right)
, \label{eq:REA6}
\end{equation}
where the RE density at $t=t_1=\Delta t_{av}$ will be given by the number of seed electrons $n_{RE} \left(t_1\right) = n_{seed} \left( t_1 \right)$, which can be evaluated from Eq. (\ref{eq:REA1}). The evolution of the number density of the first generation of secondary REs can then be expressed as:
\begin{equation}
n^{(1)}_{sec} \left( t \right) = \int d^3 p f^{(1)}_{(sec)} \left( p, \xi \right) P \left( p, \xi, \tau + t_1 ; t_{final} \right)
, \label{eq:REA7}
\end{equation}
where the RPF is now evaluated at a time $\tau+t_1 = t_{final} - \left( t - t_1 \right)$ since the first generation of secondaries is born at $t=t_1=\Delta t_{av}$. Substituting Eq. (\ref{eq:REA6}) into (\ref{eq:REA7}) yields an explicit expression for the time evolution of the number density of the first generation of secondary electrons, i.e.
\begin{equation}
n^{(1)}_{sec} \left( t \right) = 2\pi \Delta t_{av} n_e n_{RE} \left( t_1 \right)cr^2_e \int dp \frac{v}{\left( \gamma - 1 \right)^2} P \left( p, \xi_1, \tau + t_1 ; t_{final} \right)
. \label{eq:REA8}
\end{equation}
Similarly, the momentum space distribution of the second generation of secondaries is given by:
\begin{equation}
f^{(2)}_{sec} \left( p, \xi \right) = \Delta t_{av} n_e n_{RE} \left( t_2 \right) c r^2_e \frac{v}{\gamma^2-1} \frac{1}{\left( \gamma-1 \right)^2} \delta \left( \xi - \xi_1 \right)
, \label{eq:REA8sub1}
\end{equation}
where $n_{RE} \left( t_2 \right) = n_{seed} \left( t_2 \right) +
n^{(1)}_{sec} \left( t_2 \right)$.
An explicit expression for the next
generation of secondary electrons can then be written as:
\begin{align}
n^{(2)}_{sec} \left( t \right) &= \int d^3 p f^{(2)}_{(sec)} \left( p, \xi \right) P \left( p, \xi, \tau + t_2 \right) \nonumber \\
 &= 2\pi \Delta t_{av} n_e n_{RE} \left( t_2 \right)cr^2_e \int dp \frac{v}{\left( \gamma - 1 \right)^2} P \left( p, \xi_1, \tau + t_2 ; t_{final} \right)
. \label{eq:REA9}
\end{align}
where we have used Eq. (\ref{eq:REA8sub1}) in the second line. The ith generation of secondaries is then given by:
\begin{subequations}
\label{eq:REA10}
\begin{equation}
n^{(i)}_{sec} \left( t \right) = 2\pi \Delta t_{av} n_e n_{RE} \left( t_i \right)cr^2_e \int dp \frac{v}{\left( \gamma - 1 \right)^2} P \left( p, \xi_1, \tau + t_i ; t_{final} \right)
. \label{eq:REA10a}
\end{equation}
and the density of REs is
\begin{equation}
n_{RE} \left( t_i\right) = n_{seed} \left( t_i\right) + \sum^{i-1}_{j=1} n^{(j)}_{sec} \left( t_i \right)
. \label{eq:REA10c}
\end{equation}
\end{subequations}
By evaluating each successive generation of secondaries, together with the initial evolution of the seed RE population, this will enable the time history of an avalanching RE distribution to be determined. We note that while the present adjoint formulation only predicts the number density of REs, the RPF incorporates fully kinetic physics into the evolution of the RE population. The main approximation is due to the use of a Rosenbluth-Putvinski source [Eq. (\ref{eq:REA4})], which does not account for the energy or pitch distribution of the RE population.

\section{\label{sec:PINN}Physics-informed Neural Networks}

While a range of approaches to the numerical solution of Eq. (\ref{eq:TDRP9a}) could be applied, we will utilize a PINN~\cite{raissi2019physics, karniadakis2021physics}  in the present manuscript. In so doing, this will allow us to identify the parametric solution to Eq. (\ref{eq:TDRP9a}) such that once trained, we will be able to evaluate the RPF for a range of physics parameters. The specifics of our implementation of a PINN can be found in Ref. \cite{mcdevittpart12024}, however, here we will provide the essential formulae. In order to enforce several properties of the solution as hard constraints, we will introduce a physics layer of the form:
\begin{align}
P^\prime \left( p, \xi, t ; \bm{\lambda}\right) &= P_{term} \left( p \right) + \left( \frac{p-p_{min}}{p_{max}-p_{min}} \right) \tanh \left( t_{final}-t \right) P_{NN} \left( p, \xi, t ; \bm{\lambda} \right)
, \label{eq:PRED1}
\end{align}
where $\bm{\lambda}$ is a vector containing the physics parameters $\left( E_\Vert, Z_{eff}, \alpha \right)$ and the terminal condition is defined by:
\begin{equation}
P_{term} \left( p \right) \equiv \frac{1}{\Delta P} \tanh \left( \frac{p-p_{RE}}{\Delta p} \right)
, \label{eq:PRED2}
\end{equation}
where the final RPF is computed by passing Eq. (\ref{eq:PRED1}) through
\begin{equation}
P \left( p, \xi, t ; \bm{\lambda} \right) = \frac{1}{2} \left\{ 1 +  \tanh \left[ P^\prime \left( p, \xi, t ; \bm{\lambda} \right) \right] \right\}
. \label{eq:PRED6}
\end{equation}
Here, $P_{NN}$ represents the output of the hidden layers of the neural network (NN), $\Delta p/p_{max}$ and $\Delta P$ are hyperparameters whose value should be set to be much less than one, and $\bm{\lambda}$ is a vector representing the physics parameters, which are given by $\bm{\lambda} = \left( E_\Vert, Z_{eff}, \alpha \right)$. It can be verified that for small $\Delta p/p_{max}$ and $\Delta P$, the physics layer enforces (i) the terminal condition given by Eq. (\ref{eq:RPF1a}), (ii) lower momentum boundary condition [Eq. (\ref{eq:RPF1b})], and (iii) constrains the RPF to have a range between zero and one. With this physics layer, only the residual of the PDE and the boundary condition on the upper momentum boundary will be included in the loss function, such that we will minimize a loss of the form:
\begin{equation}
\text{loss} = \frac{w_{PDE}}{N_{PDE}} \sum^{N_{PDE}}_i \left[ G \left( p \right) \left( \frac{p^2_i}{1+p^2_i} \right) \mathcal{R} \left( p_i,\xi_i,t_i ; \bm{\lambda}_i\right) \right]^2 + \frac{1}{N_{bdy}} \sum^{N_{bdy}}_i \left[ P_i - P \left( p_i,\xi_i,t_i ; \bm{\lambda}_i\right) \right]^2
, \label{eq:PRED7}
\end{equation}
where
\begin{equation}
G \left( p \right) = 1 - \exp \left[ -\frac{\left( p_{max}-p \right)^2}{\Delta p_{max}^2} \right]
, \label{eq:PRED8}
\end{equation}
where $N_{bdy}$ is the number of points on the upper momentum boundary that will be sampled, and $\mathcal{R}$ is the residual of Eq. (\ref{eq:TDRP9a}). The role of the factors $G \left( p \right)$ and $p^2_i/ \left( 1+p^2_i\right)$ are to remove large contributions to the residual that emerge at either high or low momenta boundaries, as discussed further in Ref. \cite{mcdevittpart12024}. The Python script used for training the PINN is written using the DeepXDE library \cite{lu2021deepxde} with TensorFlow~\cite{abadi2016tensorflow} as the backend, and will be made available upon acceptance at https://github.com/cmcdevitt2/RunAwayPINNs.


\section{\label{sec:RDE}Temporal Evolution of an Avalanching Runaway Electron Population}

\begin{figure}
\begin{centering}
\subfigure[]{\includegraphics[scale=0.5]{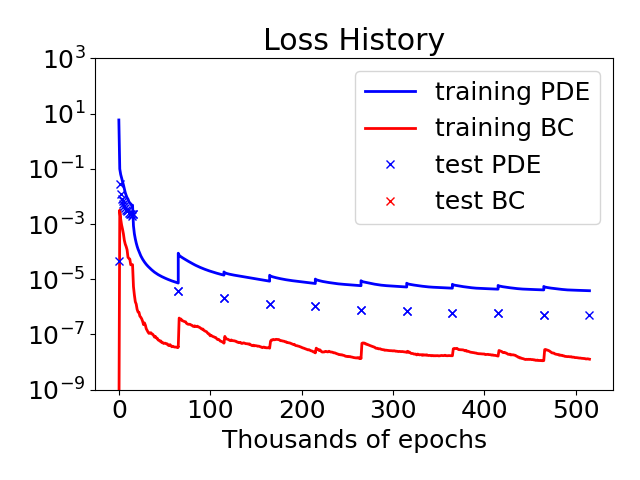}}
\subfigure[]{\includegraphics[scale=0.5]{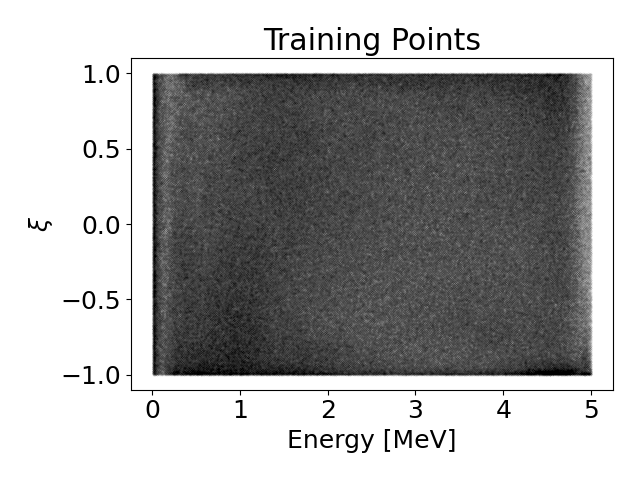}}
\par\end{centering}
\caption{Training and test loss history. The solid blue curve indicates the training loss of the PDE, the solid red curve indicates the training loss for the boundary condition, and the `x' markers indicate the test loss (the test and training losses are the same for the boundary term). 3,000,000 training and testing points are used. The training points initially obey a Hammersley sequence, with the test distribution being uniform random. The training points are resampled using the residual based adaptive distribution described in Ref. \cite{wu2023comprehensive} every 50,000 epochs of L-BFGS.}
\label{fig:TER1}
\end{figure}


This section will utilize the algorithm described in Sec. \ref{sec:REA} above to describe the temporal evolution of a RE population including large-angle collisions.
We will assume a seed electron population defined by:
\begin{equation}
f^{(init)}_e \left( p, \xi \right) \propto \exp \left[ -\frac{\left( p-p_0\right)^2}{\Delta p^2} -\frac{\left( \xi - \xi_0 \right)^2}{\Delta \xi^2} \right]
, \label{eq:REDR2}
\end{equation}
where the evolution of the number density of seed electrons will be evolved using Eq. (\ref{eq:REA1}), with each generation of secondary electrons described by Eq. (\ref{eq:REA10}). The momentum above which an electron will be counted as a RE is taken to be $p_{RE}=p_{max}/4$, which corresponds to an energy of roughly one MeV. We have set the hyperparameters to $\Delta P = 0.15$, $\Delta p = 0.1 p_{max}$, $\Delta p_{max}=0.05 p_{max}$ and $w_{PDE} = 10$ when training the model. The resulting loss history for the PINN is shown in Fig. \ref{fig:TER1}. Here we have trained the PINN for electric fields in the range $E_\Vert \in \left(0, 3 \right)$, effective charges $Z_{eff} \in \left(1, 5 \right)$, and synchrotron radiation $\alpha \in \left( 0, 0.2\right)$, a somewhat broader range of parameters compared to Ref. \cite{mcdevittpart12024}, though at the cost of a much larger number of training points. A residual based adaptive sampling scheme is employed~\cite{wu2023comprehensive}, leading to periodic spikes in the training loss. 

\begin{figure}
\begin{centering}
\subfigure[]{\includegraphics[scale=0.5]{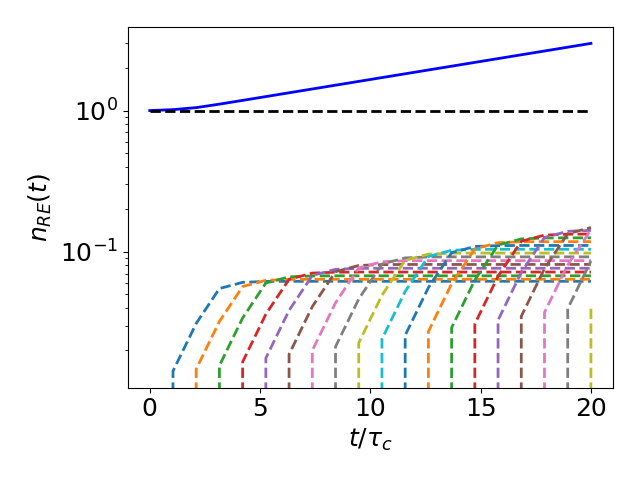}}
\subfigure[]{\includegraphics[scale=0.5]{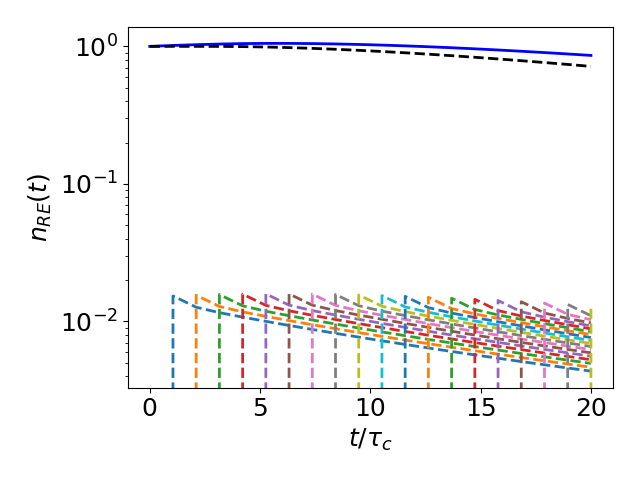}}
\par\end{centering}
\caption{Runaway electron evolution over time for two different values of the electric field. The number of total REs is shown by the blue curves, the seed population is indicated by the black dashed curve, and each generation of secondaries are indicated by the multicolor dashed curves. Panel (a) is for $E_\Vert = 3$ (above marginality) whereas panel (b) is for $E_\Vert = 1.5$ (below marginality). The other parameters are given by $Z_{eff}=1$, $\alpha = 0.1$ and $\ln \Lambda = 15$, with 19 generations of secondaries included (i.e. twenty time steps). The seed electron distribution was taken to be Eq. (\ref{eq:REDR2}) with $\left( p_0=3p_{max}/4,\Delta p = p_{max}/10,\xi_0 = -1, \Delta \xi = 0.1\right)$.}
\label{fig:AAS1}
\end{figure}

Two example time histories are shown in Fig. \ref{fig:AAS1}. Here, the time evolution of the total number of REs (solid blue curve), initial seed population (dashed black curve) and several generations of secondary REs (multicolored dashed curves) are shown. For panel (a) of Fig. \ref{fig:AAS1}, the electric field was chosen to be above threshold ($E_\Vert=3$), with the total number of REs increasing exponentially (solid blue curve). Considering each generation, the seed electron population is normalized to have a value of one at $t=0$, where its value remains largely unchanged since $E_\Vert > E_{av}$ ($E_{av}\approx 1.8$ for these parameters). For each avalanche time step, a small number of secondary electrons are born, where each generation grows rapidly until plateauing later in time. The initial rapid increase in the secondary electron population is due to the energy and pitch distribution of secondaries immediately after they are born obeying Eq. (\ref{eq:REA4}), where the majority of these electrons will have energies below the one MeV threshold energy for an electron to be counted as a RE. For cases above threshold, a large number of the secondary electrons will be quickly accelerated above one MeV, resulting in an increase in the total number of REs. This process is repeated for each generation of secondaries, with each generation having a larger magnitude due to the number of REs increasing according to Eq. (\ref{eq:REA10c}). Summing together the number of primary electrons with all generations of secondary electrons results in an exponentially increasing number of REs, as indicated by the solid blue curve in Fig. \ref{fig:AAS1}.
Turning to Fig. \ref{fig:AAS1}(b), here the electric field was chosen to be below threshold ($E_\Vert = 1.5$), with the total number of REs initially increasing slightly before decaying slowly in time. Considering first the seed runaway population, the number of seed electrons initially remains approximately constant, but then begins to decay at a later time. The finite number of seed electrons leads to the generation of secondary electrons, and hence an initial increase in the total number of REs. However, since each secondary generation decays exponentially in time this does not lead to the long term growth of the overall population. At late times the total number of REs decay exponentially with a rate similar, though not identical to, the decay of the seed RE population.

\begin{figure}
\begin{centering}
\subfigure[]{\includegraphics[scale=0.33]{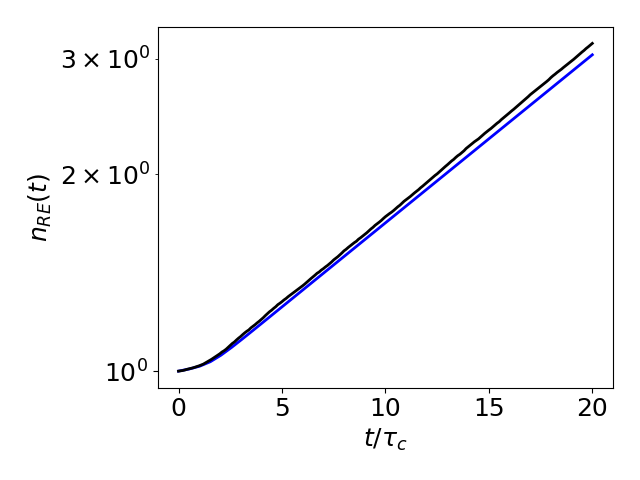}}
\subfigure[]{\includegraphics[scale=0.33]{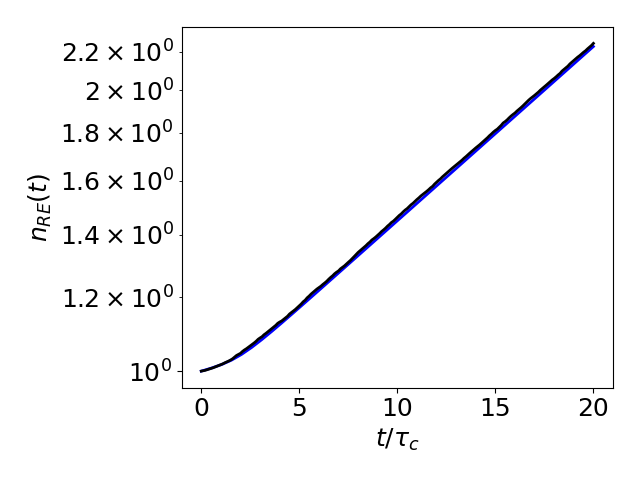}}
\subfigure[]{\includegraphics[scale=0.33]{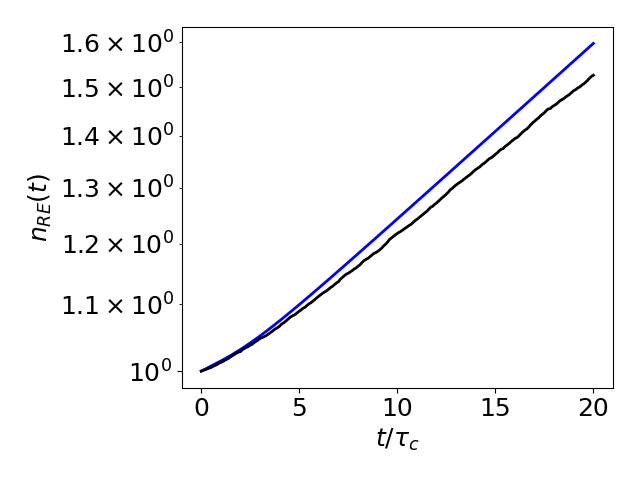}}
\subfigure[]{\includegraphics[scale=0.33]{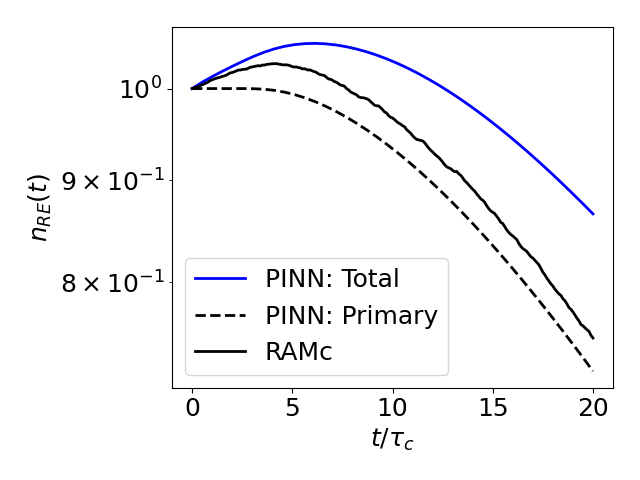}}
\subfigure[]{\includegraphics[scale=0.33]{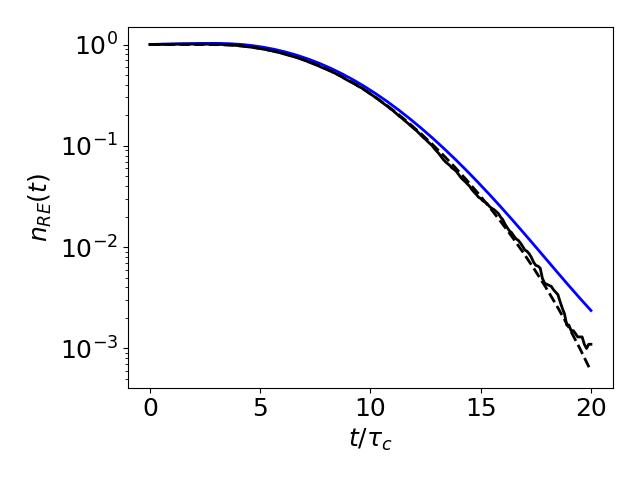}}
\subfigure[]{\includegraphics[scale=0.33]{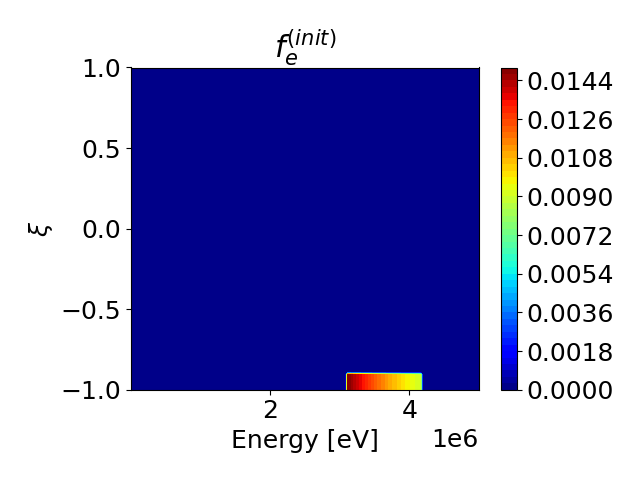}}
\par\end{centering}
\caption{Comparison of RE density evolution with the Monte Carlo code RAMc for electric fields $E_\Vert=3$ [panel (a)], $E_\Vert=2.5$ [panel (b)], $E_\Vert=2$ [panel (c)], $E_\Vert=1.5$ [panel (d)], and $E_\Vert=1$ [panel (e)]. The initial electron distribution assumed for both RAMc and the PINN is shown in panel (f), where a rectangular block of electrons were inserted. The other parameters were taken to be $Z_{eff}=1$, $\alpha = 0.1$ and $\ln \Lambda = 15$.}
\label{fig:AAS2sub1}
\end{figure}

The RE density histories for several different electric fields are shown in Fig. \ref{fig:AAS2sub1}, where the PINN's predictions (blue curves) are compared with the RE solver RAMc (black curves). RAMc is a particle based RE solver~\cite{mcdevitt2019avalanche} that evolves guiding center orbits together with small and large angle collisions, as well as synchrotron radiation. For the present case the electrons were initialized on the magnetic axis for a large ITER like device, such that neoclassical effects arising from magnetic trapping~\cite{Rosenbluth:1997, Chiu:1998, mcdevitt2019runaway, arnaud2024impact} along with spatial transport~\cite{mcdevitt2019spatial} are negligible. This will allow for comparisons with the present formulation, which does not account for toroidal geometry or spatial transport. Furthermore, RAMc uses a M\o ller source [Eqs. (\ref{eq:REA2sub1}) and (\ref{eq:REA3})] to describe secondary electron generation, a higher physics fidelity treatment of large-angle collisions compared to the Rosenbluth-Putvniski source used in the adjoint-deep learning approach. The present comparison will thus test the accuracy of the adjoint-deep learning framework along with regimes where a Rosenbluth-Putvinski source provides an adequate approximation to the large-angle collision operator. In all cases, the electron distribution was initialized to be a block of particles at high energies with pitch near $\xi \approx -1$
[see Fig. \ref{fig:AAS2sub1}(f)]
to be consistent with the initial RE distribution implemented in the RAMc code. From Figs. \ref{fig:AAS2sub1}(a) and (b) it is evident that when well above threshold the two approaches are in good agreement. However, when approaching threshold [panels (c) and (d) of Fig. \ref{fig:AAS2sub1}], more substantial differences between the two approaches are evident, with the PINN systematically over predicting the RE density. The cause of this discrepancy can be linked to the Rosenbluth-Putvinski secondary source, which assumes the primary electrons to have asymptotically large energy. While very high energy primary electrons will be present when the system is above marginality, no such electrons are expected to be present when at or below marginality. As a result, the Rosenbluth-Putvinski source will prove inaccurate when near threshold. Comparing the time evolution of the primary electron population [dashed curves in Figs. \ref{fig:AAS2sub1}(d) and (e)] with the result from RAMc, it is evident that the decay rate of the \emph{primary} electron population is in better agreement with the results of RAMc. The reason for this is that RAMc employs a M\o ller secondary source term that accounts for the energy and pitch of the primary electrons. When below threshold, the energy of the primary distribution will be modest, such that the secondary electrons generated by the M\o ller source will have little impact on the RE number density, with the majority of them being born with too low an energy to run away. Thus, 
as the system drops further below marginality, neglecting the Rosenbluth-Putvinski source in the PINN leads to an increasingly good approximation to the RE density evolution.

\section{\label{sec:PVR}Parametric Variation of Runaway Electron Avalanche Growth and Decay Rate}

\begin{figure}
\begin{centering}
\subfigure[]{\includegraphics[scale=0.33]{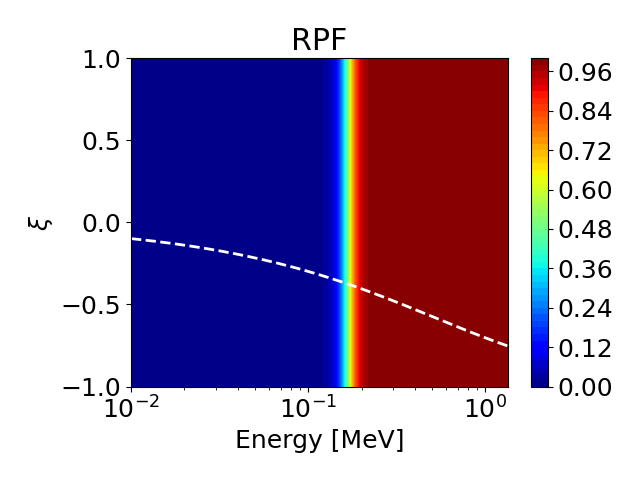}}
\subfigure[]{\includegraphics[scale=0.33]{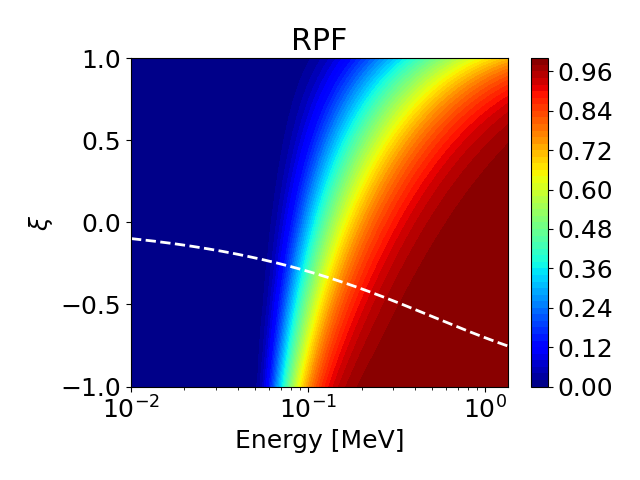}}
\subfigure[]{\includegraphics[scale=0.33]{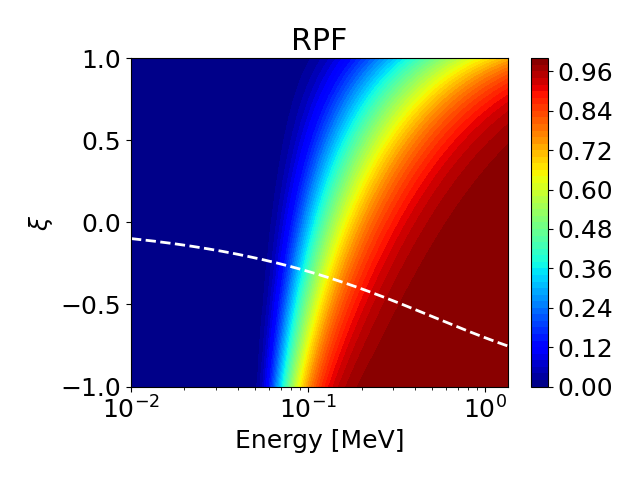}}
\subfigure[]{\includegraphics[scale=0.33]{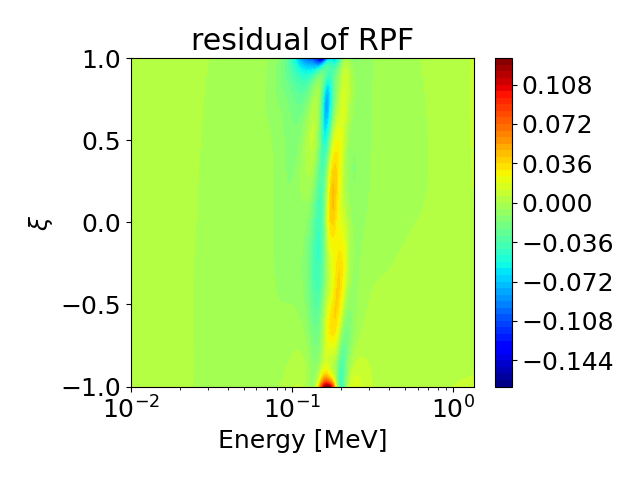}}
\subfigure[]{\includegraphics[scale=0.33]{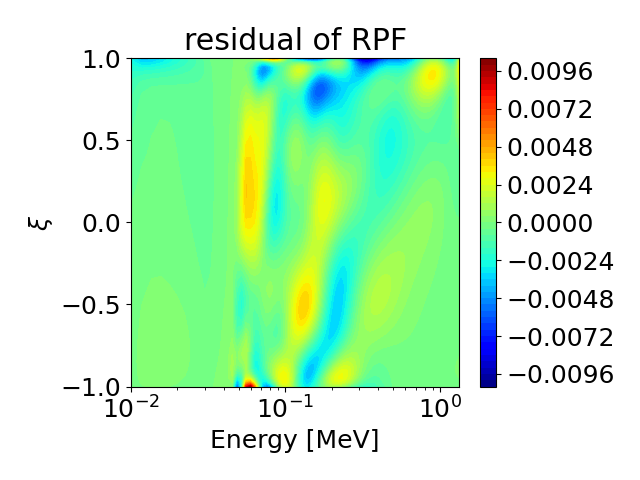}}
\subfigure[]{\includegraphics[scale=0.33]{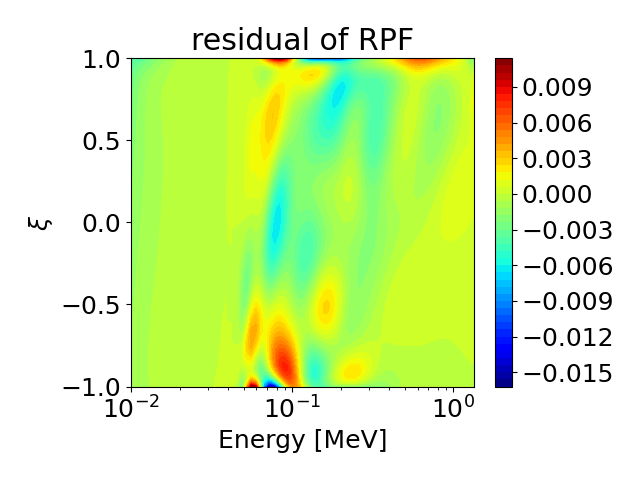}}
\par\end{centering}
\caption{Time slices of the RPF evolution (top row) along with the associated residual (bottom row) for $E_\Vert = 8$ and $p_{RE}=\bar{p}_{max}/4$. The bottom row indicates the residual to Eq. (\ref{eq:TDRP9a}) multiplied by $p^2/\left( 1+p^2\right)$, such that the low energy divergence is removed. The first column indicates the terminal condition at $t\approx 2.19$, the second column $t\approx 1.65$ and the last column $t=0$. The dashed white curves are the location of secondary injection $\xi_1$ defined by Eq. (\ref{eq:REA5}). The other parameters are given by $Z_{eff}=5$ and $\alpha = 0.1$, and we chose $t_{max}=20$, $N_p=10$, $N_{av}=0.5$, along with a $p_{max}$ consistent with $10\;\text{MeV}$.}
\label{fig:PVR3}
\end{figure}

Our aim in the present section will be to train the PINN across a broader range of physics parameters. The treatment of a broader range of parameters will pose a substantial challenge to training an accurate PINN. Specifically, as the size of the parameter space is increased, the number of training points required to densely sample the space will grow substantially, thus requiring a longer period of offline training along with increasing memory usage. In addition, the location of the transition region of the RPF (i.e. where the RPF transitions between zero and one), and the time to reach a steady state solution, will traverse a broad range of energy and time scales further complicating capturing the RPF with a single PINN. To accommodate both of these challenges we will introduce a normalized time and energy coordinate that will allow the PINN to adapt the time period and the energy range that must be captured in order to accurately infer the RPF evolution. To normalize the time period over which the RPF is evolved we will utilize the Rosenbluth-Putvinski growth rate as a rough estimate of the rate that the RE density changes, i.e.~\cite{Rosenbluth:1997}
\begin{equation}
\tau_c  \gamma_{RP}= \frac{1}{\ln \Lambda} \sqrt{\frac{\pi}{3\left( Z_{eff}+5\right)}} \left( E_\Vert - 1 \right) 
, \label{eq:PVR1}
\end{equation}
where the parallel electric field is normalized to the Connor-Hastie threshold $E_c$.
The time period over which the RPF will be evolved, denoted by $\bar{t}_{final}$, will be taken to have the form:
\begin{equation}
\bar{t}_{final} = \frac{t_{final}}{\sqrt{1+ \gamma^2_{av}t^2_{final}/N^2_{av}}}
, \label{eq:PVR2}
\end{equation}
where $N_{av}$ is the number of avalanche timescales that will be simulated and $t_{final}$ provides an upper bound on the time period of the simulation. It may be readily verified that near threshold where $t_{final}\gamma_{av}/N_{av} \ll 1$, Eq. (\ref{eq:PVR2}) asymptotes to $\bar{t}_{final} \approx t_{final}$, whereas when far from threshold where $t_{max}\gamma_{av} / N_{av} \gg 1$, $\bar{t}_{final}$ asymptotes to $\bar{t}_{final} \approx N_{av} / \gamma_{av}$. Equation (\ref{eq:PVR2}) thus allows for relatively long simulations to be carried out when the system is near threshold, but substantially shrinks the time domain for scenarios far above threshold where the RPF quickly reaches a quasi steady state [see Fig. \ref{fig:PVR3} below].

The upper bound in the momentum domain will be chosen to be larger than the momentum where the electric field acceleration and drag balance for an electron with $\xi=-1$, which defines a critical momentum given by $p_{crit} = 1/ \sqrt{E_\Vert - 1}$. To avoid the singularity at $E_\Vert = 1$, we will approximate this critical momentum by $p_{crit} \approx 1/ \sqrt{E_\Vert+\delta}$, where $\delta$ is a small value that removes the divergence at $E_\Vert = 0$, taken to be $\delta = 0.01$ in the present study. The upper momentum bound will then be taken to have the form
\begin{equation}
\bar{p}_{max} = \frac{p_{max}}{\sqrt{1+p^2_{max}/\left( N_p p_{crit}\right)^2}}
, \label{eq:PVR3}
\end{equation}
where $N_p$ is a factor that sets how far above $p_{crit}$ the high momentum boundary is set, and $p_{max}$ limits the upper momentum boundary for the case of a weak electric field. For $p^2_{max}/\left( N_p p_{crit}\right)^2 \ll 1$ (i.e. a weak electric field), the upper momentum boundary is approximately given by $\bar{p}_{max} \approx p_{max}$, whereas for a strong electric field $p^2_{max}/\left( N_p p_{crit}\right)^2 \gg 1$, Eq. (\ref{eq:PVR3}) asymptotes to $\bar{p}_{max} \approx N_p p_{crit}$. An upper momentum bound given by Eq. (\ref{eq:PVR3}) thus allows a high momentum boundary of roughly $p_{max}$ to be used when near or below marginality, but will substantially shrink the momentum domain for large electric fields where the transition region is located at low momenta.

\begin{figure}
\begin{centering}
\subfigure[]{\includegraphics[scale=0.5]{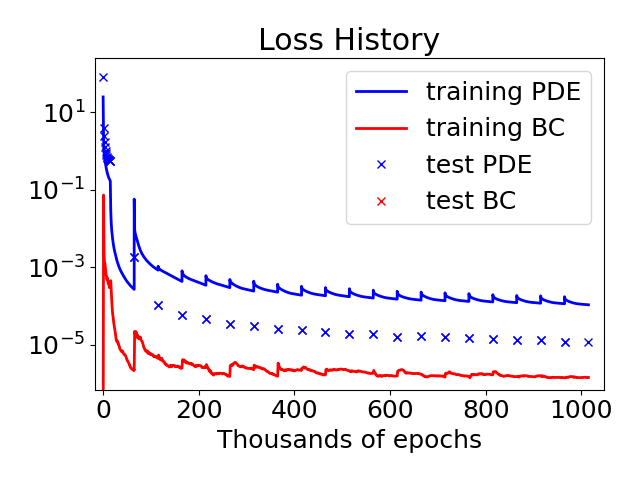}}
\subfigure[]{\includegraphics[scale=0.5]{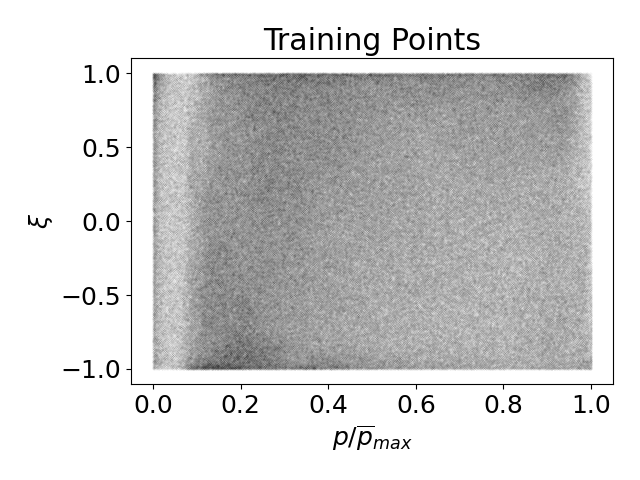}}
\par\end{centering}
\caption{(a) Training and test loss history. The solid blue curve indicates the training loss of the PDE, the solid red curve indicates the training loss for the boundary condition, and the `x' markers indicate the test loss (the test and training losses are the same for the boundary term). 1,000,000 training and test points were used. (b) Final training point distribution.}
\label{fig:PVR1}
\end{figure}

Utilizing these adaptive time and momentum ranges, the loss history and final training point distribution are shown in Fig. \ref{fig:PVR1}. Here we have trained the PINN to learn the RPF for electric fields in the range $E_\Vert \in \left( 0, 10\right)$, $Z_{eff} \in \left( 1, 10\right)$ and $\alpha \in \left( 0,0.2\right)$ for the hyperparameters $\Delta P = 0.2$, $\Delta \bar{p} = 0.1 p_{max}$, $\Delta p_{max}=0.05 \bar{p}_{max}$ and $w_{PDE} = 10$. The time and momentum domains are defined using Eqs. (\ref{eq:PVR1})-(\ref{eq:PVR3}) with the values $t_{max}=20$, $N_{av} = 1/2$, $N_p = 10$, and a $p_{max}$ is chosen consistent with $10\;\text{MeV}$. The low momentum boundary is taken to be consistent with $10\;\text{keV}$ and $p_{RE}$ is set to be $p_{RE} = \bar{p}_{max} / 4$.  We also slightly modified the loss function to
\begin{equation}
\text{loss} = \frac{w_{PDE}}{N_{PDE}} \sum^{N_{PDE}}_i \left[ G \left( p \right) \frac{E_{avg}}{1+E_\Vert} \left( \frac{p^2_i}{1+p^2_i} \right) \mathcal{R} \left( p_i,\xi_i,t_i ; \bm{\lambda}_i\right) \right]^2 + \frac{1}{N_{bdy}} \sum^{N_{bdy}}_i \left[ P_i - P \left( p_i,\xi_i,t_i ; \bm{\lambda}_i\right) \right]^2
, \label{eq:PRR4}
\end{equation}
where we have included the additional factor $E_{avg}/ \left( 1+E_\Vert \right)$ in front of the residual, where $E_{avg} = \left( E_{min} + E_{max} \right)/2$, due to the magnitude of the residual of Eq. (\ref{eq:TDRP9a}) increasing with electric field strength. By dividing by the electric field in the loss, this prevents the training of the PINN from being unduly biased toward cases with large electric fields. A residual based adaptive training point sampling scheme~\cite{wu2023comprehensive} has been adopted, where from Fig. \ref{fig:PVR1}(b) it is evident that the density of training points is highest near $p/\bar{p}_{max} \approx 0.2$ and at the high momentum boundary near $\xi \approx \pm 1$. The concentration of training points near $p/\bar{p}_{max} \approx 0.2$ is due to the value of $p_{RE}$ being chosen to be $p_{RE}/\bar{p}_{max} = 0.25$, whereas the concentration of points near the upper momentum boundary with $\xi \approx \pm 1$ often contains the maximum value of the RPF when below threshold (see the second column of Fig. \ref{fig:PVR2} below).

\begin{figure}
\begin{centering}
\subfigure[]{\includegraphics[scale=0.33]{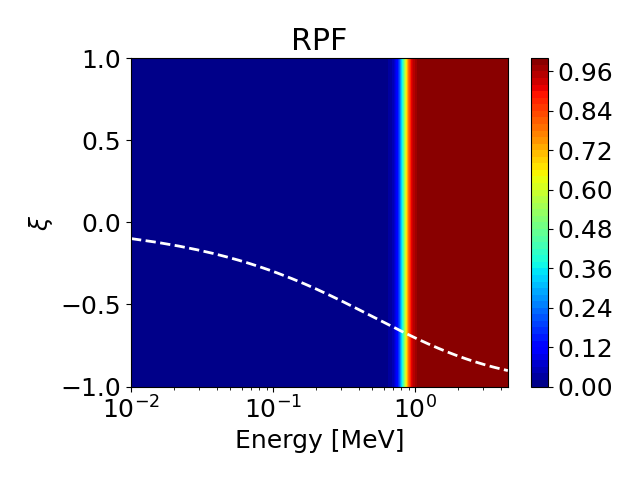}}
\subfigure[]{\includegraphics[scale=0.33]{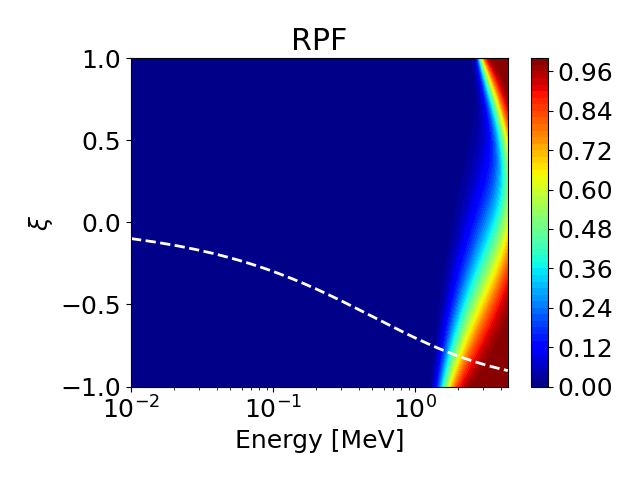}}
\subfigure[]{\includegraphics[scale=0.33]{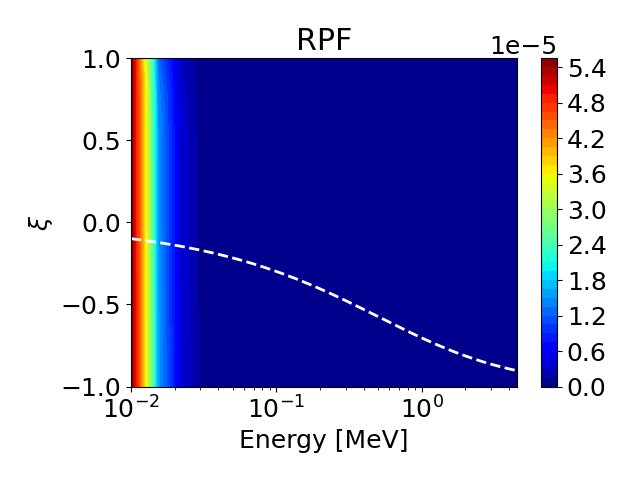}}
\subfigure[]{\includegraphics[scale=0.33]{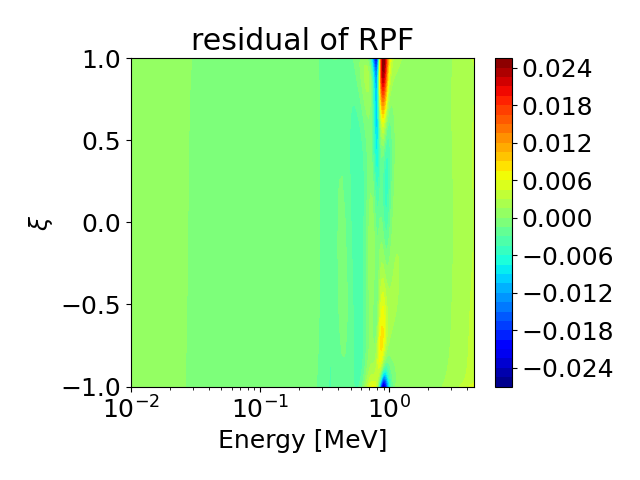}}
\subfigure[]{\includegraphics[scale=0.33]{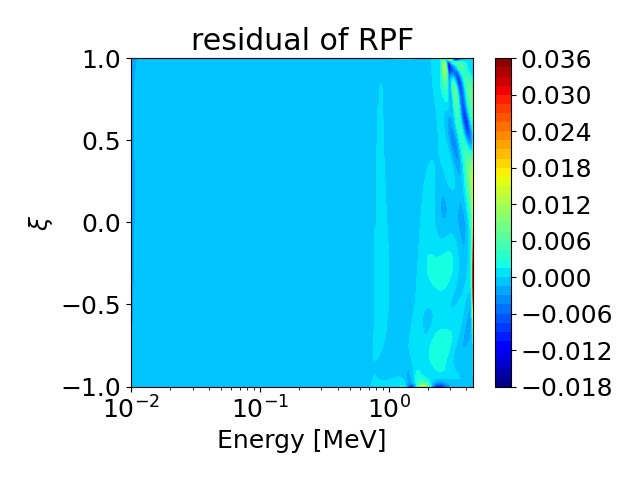}}
\subfigure[]{\includegraphics[scale=0.33]{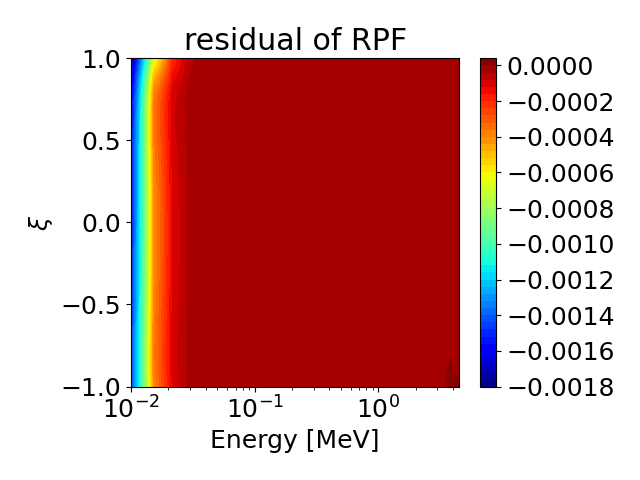}}
\par\end{centering}
\caption{Time slices of the RPF evolution (top row) along with associated residuals (bottom row) for $E_\Vert = 0.8$ and $p_{RE}=\bar{p}_{max}/4$. The bottom row indicates the residual to Eq. (\ref{eq:TDRP9a}) multiplied by $p^2/\left( 1+p^2\right)$, such that the low energy divergence is removed. The first column indicates the terminal condition at $t\approx 19.36$, the second column $t\approx 17.43$ and the last column $t=0$. The dashed white curves are the location of secondary injection $\xi_1$ defined by Eq. (\ref{eq:REA5}). The other parameters are given by $Z_{eff}=5$ and $\alpha = 0.1$, and we chose $t_{max}=10$, $N_p=10$, $N_{av}=0.5$, along with a $p_{max}$ consistent with $10\;\text{MeV}$}
\label{fig:PVR2}
\end{figure}

Two example RPFs for both weak and large electric fields are shown in Figs. \ref{fig:PVR2} and \ref{fig:PVR3}, respectively. For the case of a weak electric field ($E_\Vert=0.8$, $Z_{eff}=5$, $\alpha = 0.1$), a large energy domain is used, with an upper bound of $\approx 4.5 \;\text{MeV}$ along with an integration time of nearly twenty $\tau_c$, such that the relatively slow evolution of the RPF can be captured. Here, the terminal condition is chosen such that $p_{RE} = \bar{p}_{max}/4$, which for the weak electric field case corresponds to an energy slightly below $1\;\text{MeV}$. As time evolves the transition region shifts to higher energy, with maximums near $\xi \approx \pm 1$ at $t\approx 17.43$ (middle column in Fig. \ref{fig:PVR2}). Here, these maximums at high energy are due to synchrotron radiation preferentially slowing down electrons with large quantities of perpendicular energy (i.e. with $\xi \approx 0$). By $t=0$ the RPF is nearly zero everywhere, except near the low energy boundary where it has a magnitude of $\sim 10^{-6}$. The reason the RPF does not vanish identically at the low energy boundary is due to the physics layer defined by Eqs. (\ref{eq:PRED1})-(\ref{eq:PRED6}) not precisely vanishing at $p=p_{min}$. In particular, at $p=p_{min}$, the RPF reduces to [see Eq. (\ref{eq:PRED6})]
\[
P\left( p_{min}\right) = 0.5 \left\{ 1+\tanh \left[ P_{term} \left( p_{min}\right) \right] \right\}
,
\]
which for $\Delta P = 0.2$ and $\Delta p = 0.1\bar{p}_{max}$ does not precisely vanish. This small inaccuracy will have a negligible impact for parameters far above threshold, and will only impact predictions on the PINN at late times when well below marginality. Furthermore, since the spurious contribution to the RPF is located at the low energy boundary, it can be removed via a post processing routine as discussed below.

Considering now a case well above marginality ($E_\Vert=8$, $Z_{eff}=5$, $\alpha = 0.1$, see Fig. \ref{fig:PVR3}), the RPF quickly reaches a steady state, with the RPF at $t\approx 1.65$ having nearly the same form as at $t=0$, with the total time of the simulation being $2.19$. The steady state solution has a contour of roughly one half at an energy of $60\;\text{keV}$, suggesting that the maximum energy used in this case of $\approx 1\;\text{MeV}$ is sufficient to capture the critical portion of the RPF. 

\begin{figure}
\begin{centering}
\includegraphics[scale=0.5]{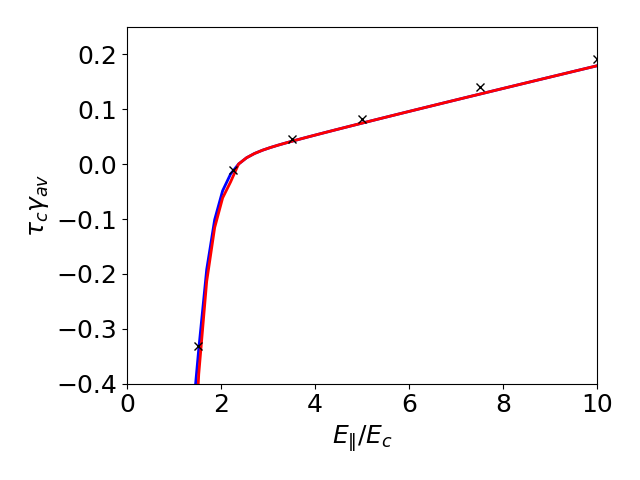}
\par\end{centering}
\caption{Comparison of avalanche growth rates between the Monte Carlo code RAMc with a M\o ller secondary source and predictions of the PINN. The solid blue curve includes large-angle collisions, whereas the solid red curve includes large-angle collisions when above threshold, but uses the primary decay rate when below threshold, and the black `x' markers are values from RAMc. The other parameters were $Z_{eff}=5$, $\alpha = 0.1$, and $\ln \Lambda = 16.15$. The initial electron distribution was defined by Eq. (\ref{eq:REDR2}) with $\left( p_0 = 3\bar{p}_{max}/4,\Delta p = \bar{p}_{max}/10,\xi_0 = -1, \Delta \xi = 0.1\right)$.}
\label{fig:PVR4}
\end{figure}

Using the PINN described in Figs. \ref{fig:PVR1}-\ref{fig:PVR3} we will be interested in comparing predictions of the RE avalanche growth or decay rate with Monte Carlo simulations carried out with RAMc. Before carrying out this comparison we note that the use of a Rosenbluth-Putvinski source to describe large-angle collisions introduces an ambiguity in the evaluation of the avalanche growth rate. In particular, since the Rosenbluth-Putvinski source only depends on the number of REs, and not the energy distribution, it is necessary to introduce a criterion for classifying which electrons should be counted as REs. In the present formulation the quantity $p_{RE}$ defined by Eq. (\ref{eq:RPF1}) acts as the effective definition of which electrons are counted as REs. In Fig. \ref{fig:PVR4} this value is roughly $2.5p_{crit}$ for the hyperparameters selected, where $p_{crit}$ is the critical energy to run away given by $p_{crit} \approx 1/ E_\Vert$. This definition ensures that only electrons located substantially above the critical energy to run away are included in the large-angle collision operator. 

An additional subtly that arises when using an adaptive energy region, is that for the case of a large electric field, the modest value of the upper momentum boundary results in the integral used to evaluate the number of secondary electrons [Eq. (\ref{eq:REA10a})] being prematurely cutoff. This can result in underestimating the number of secondary electrons generated. To account for this we extended the energy integral in Eq. (\ref{eq:REA10a}) to infinity, where we take the value of the RPF to be $P \left( \bar{p}_{max},\xi_1 \right)$ for the extended region of the integral. For cases well above threshold, we will have $P \left( \bar{p}_{max},\xi_1 \right) \approx 1$ (see the dashed white curves in Fig. \ref{fig:PVR3}, for example), and we would expect this quantity to remain near unity for $p > \bar{p}_{max}$ had a higher upper momentum boundary been used.
In contrast, when near or below threshold, $P \left( \bar{p}_{max},\xi_1 \right)$ will no longer be unity, and will likely vary significantly at higher momentum. For this latter case, however, the high energy boundary will be located at several MeV (see Fig. \ref{fig:PVR2}, for example), where at such high energies the Rosenbluth-Putvinski source has a negligibly small value. In this limit the contribution from the high energy extension of the integral defined by Eq. (\ref{eq:REA10a}) will make a negligible contribution to the rate of RE growth or decay, such that our extension of the integral defined in Eq. (\ref{eq:REA10a}) past the $\bar{p}_{max}$ boundary will have little impact on the predicted number of secondary electrons.

Finally, as evident in Fig. \ref{fig:PVR2}(c), when well below threshold, a spurious contribution to the RPF can emerge at the low energy boundary. To remove this contribution we will note two momenta. The first is the critical momentum for an electron to overcome drag, defined by $p_c \equiv 1 / \sqrt{E_\Vert - 1}$, for $E_\Vert\leq 1$, or $p_c \to \infty$ otherwise. The second is a hybrid momentum formed by taking the weighted average of $p_{min}$ and $p_{RE}$, i.e. $p_{ref} = \left( p_{RE} + 3p_{min}\right) / 4$. To remove the spurious contribution to the RPF near $p_{min}$, we will multiply the integrand appearing in Eq. (\ref{eq:REA10a}) by a Heaviside function $\Theta \left( p - p_{c}\right)$ for $p_{c} < p_{ref}$, thus removing electrons below the critical momentum, or $\Theta \left( p - p_{ref}\right)$ otherwise. In so doing, contributions from momenta below $p_{ref}$ will be removed when near or below threshold, but when far above threshold (i.e. when $p_c < p_{ref}$), only momenta below $p_c$ are removed. Thus, in both limits the spurious contribution to the RPF near $p=p_{min}$ is removed, but when well above threshold only very small values of momenta are impacted.

\begin{figure}
\begin{centering}
\subfigure[]{\includegraphics[scale=0.5]{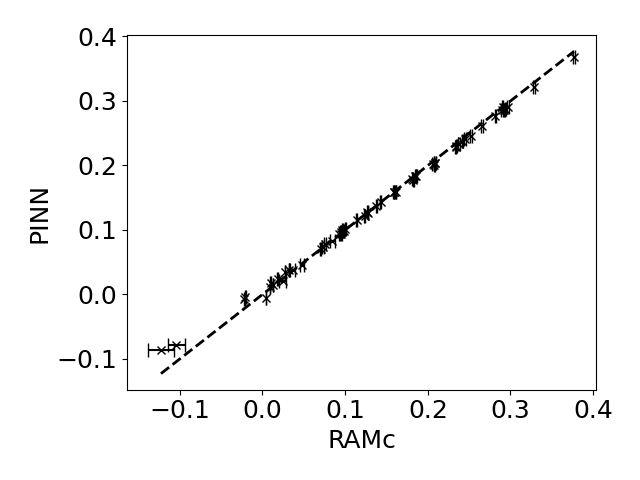}}
\subfigure[]{\includegraphics[scale=0.5]{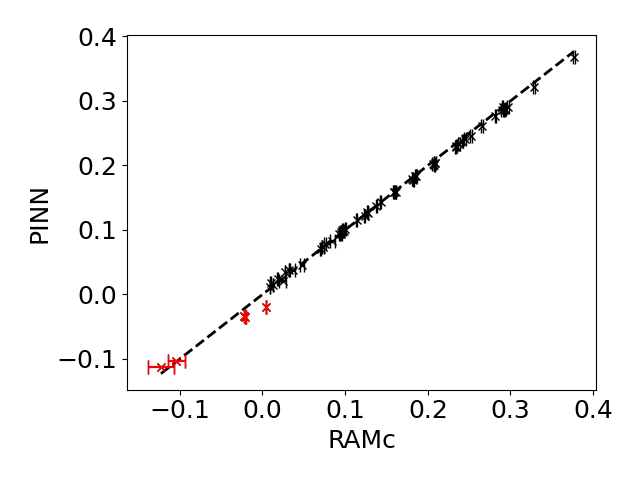}}
\par\end{centering}
\caption{Comparison of avalanche growth and decay rates between the Monte Carlo code RAMc and the PINN over a broad range of plasma conditions. Panel (a) includes a Rosenbluth-Putvinski secondary source term for all the PINN's predictions, whereas panel (b) uses the decay rate of the primary distribution when the PINN predicts RE decay. The electric field, effective charge $Z_{eff}$, and synchrotron radiation $\alpha$ were selected randomly over the region the PINN was trained. The Coulomb logarithm was taken to be $\ln \Lambda = 9.998$ in this scan. The initial electron distribution was defined by Eq. (\ref{eq:REDR2}) with $\left( p_0 = 3\bar{p}_{max}/4,\Delta p = \bar{p}_{max}/10,\xi_0 = -1, \Delta \xi = 0.1\right)$.}
\label{fig:PVR5}
\end{figure}

Using the above algorithm, the avalanche growth or decay of an initial seed RE population has been evaluated for cases both well above and below threshold, with the result compared with RAMc simulations that employ a M\o ller secondary source [see Fig. \ref{fig:PVR4}]. Good agreement is found both for a strongly decaying RE distribution along with a RE distribution that is rapidly growing. Considering a broader comparison between predictions from the PINN and RAMc, 50 values of the parameters $\left( E_\Vert, Z_{eff}, \alpha \right)$ were randomly sampled over the range of parameters the PINN was trained across. Cases that decayed too rapidly for a good exponential fit to be identified, which corresponded to cases far below threshold as discussed in Ref. \cite{mcdevittpart12024}, were discarded leaving 44 cases that will be compared with PINN predictions. The resulting comparison between the two approaches is shown in Fig. \ref{fig:PVR5}(a). Generally good agreement is evident, suggesting the PINN is able to describe the growth or decay of a RE population across a range of plasma conditions. We note that the rapid increase in the magnitude of the RE decay rate as the electric field drops below threshold (i.e. the steep drop off evident in Fig. \ref{fig:PVR4}) resulted in only a few data points that were substantially below threshold, but not so far below threshold to prevent an exponential fit, being present in our randomly sampled dataset. 

The largest disagreement between the predictions of the PINN and RAMc occurs when below threshold. As noted in Fig. \ref{fig:AAS2sub1}, and the ensuing discussion, when below threshold the Rosenbluth-Putvinski source over predicts the impact of secondary electron generation on the RE density, where a better approximation can be achieved by neglecting the secondary source and computing the decay rate of the primary distribution. This is done in Fig. \ref{fig:PVR5}(b), where the decay rate of the values (red `x' markers) that the PINN predicted were below threshold are now computed using the decay rate of the primary distribution, yielding improved agreement with predictions from RAMc, particularly for the strongest decaying cases.

\section{\label{sec:C}Conclusion}

An adjoint-deep learning framework for evaluating the time evolution of the RE density incorporating large-angle collisions was derived. By utilizing a PINN to identify the parametric solution to the adjoint problem, this enables the time evolution of the RE density to be rapidly inferred across a broad range of parameters for an arbitrary initial momentum space distribution. 
This framework was applied to evaluate time histories of the RE density, along with saturated growth and decay rates across a broad range of physics parameters. Excellent agreement was found between predictions of the adjoint-deep learning framework and a traditional RE solver for the saturated avalanche growth rate. It was found that near and below marginality, the Rosenbluth-Putvinski approximation to the secondary source term lead to quantitatively inaccurate time histories of the RE density, though the qualitative behavior was captured. Furthermore, it was found that when below marginality, the decay rate of the RE distribution is well approximated by the decay rate of the primary distribution function (i.e. neglecting large-angle collisions). We anticipate that extensions of the present framework to include a more comprehensive set of physics including time varying electric fields, temporally varying plasma compositions, along with partial screening corrections to collisional coefficients will allow the present approach to provide an efficient, yet accurate tool for evaluating saturated RE growth and decay rates, as well as describing RE dynamics during rapid variations in the background plasma. Such extensions will be the topic of future work.

\begin{acknowledgements}

This work was supported by the Department of Energy, Office of Fusion Energy Sciences at the University of Florida under awards DE-SC0024649 and DE-SC0024634, and at Los Alamos National Laboratory (LANL) under contract No. 89233218CNA000001. The authors acknowledge the University of Florida Research Computing for providing computational resources that have contributed to the research results reported in this publication. This research used resources of the National Energy Research Scientific Computing Center (NERSC), a Department of Energy Office of Science User Facility using NERSC award FES-ERCAP0028155.

\end{acknowledgements}

%
%
%
%


\begin{thebibliography}{41}
\expandafter\ifx\csname natexlab\endcsname\relax\def\natexlab#1{#1}\fi
\expandafter\ifx\csname bibnamefont\endcsname\relax
  \def\bibnamefont#1{#1}\fi
\expandafter\ifx\csname bibfnamefont\endcsname\relax
  \def\bibfnamefont#1{#1}\fi
\expandafter\ifx\csname citenamefont\endcsname\relax
  \def\citenamefont#1{#1}\fi
\expandafter\ifx\csname url\endcsname\relax
  \def\url#1{\texttt{#1}}\fi
\expandafter\ifx\csname urlprefix\endcsname\relax\def\urlprefix{URL }\fi
\providecommand{\bibinfo}[2]{#2}
\providecommand{\eprint}[2][]{\url{#2}}

\bibitem[{\citenamefont{McDevitt{\it~et al.}}(2024)}]{mcdevittpart12024}
\bibinfo{author}{\bibfnamefont{C.}~\bibnamefont{McDevitt{\it~et al.}}},
  \emph{\bibinfo{title}{A physics-constrained deep learning treatment of
  runaway electron dynamics}}, \bibinfo{howpublished}{Submitted to Physics of
  Plasmas} (\bibinfo{year}{2024}).

\bibitem[{\citenamefont{Kruskal and Bernstein}(1962)}]{Kruskal-Bernstein:1962}
\bibinfo{author}{\bibfnamefont{M.}~\bibnamefont{Kruskal}} \bibnamefont{and}
  \bibinfo{author}{\bibfnamefont{I.}~\bibnamefont{Bernstein}},
  \bibinfo{howpublished}{Princeton Plasma Physics Lab. Report No. Matt-Q-20
  (unpublished)} (\bibinfo{year}{1962}).

\bibitem[{\citenamefont{Connor and Hastie}(1975)}]{Connor:1975}
\bibinfo{author}{\bibfnamefont{J.}~\bibnamefont{Connor}} \bibnamefont{and}
  \bibinfo{author}{\bibfnamefont{R.}~\bibnamefont{Hastie}},
  \bibinfo{journal}{Nuclear fusion} \textbf{\bibinfo{volume}{15}},
  \bibinfo{pages}{415} (\bibinfo{year}{1975}).

\bibitem[{\citenamefont{Rosenbluth and Putvinski}(1997)}]{Rosenbluth:1997}
\bibinfo{author}{\bibfnamefont{M.}~\bibnamefont{Rosenbluth}} \bibnamefont{and}
  \bibinfo{author}{\bibfnamefont{S.}~\bibnamefont{Putvinski}},
  \bibinfo{journal}{Nuclear Fusion} \textbf{\bibinfo{volume}{37}},
  \bibinfo{pages}{1355} (\bibinfo{year}{1997}).

\bibitem[{\citenamefont{Smith and Verwichte}(2008)}]{smith2008hot}
\bibinfo{author}{\bibfnamefont{H.~M.} \bibnamefont{Smith}} \bibnamefont{and}
  \bibinfo{author}{\bibfnamefont{E.}~\bibnamefont{Verwichte}},
  \bibinfo{journal}{Physics of plasmas} \textbf{\bibinfo{volume}{15}},
  \bibinfo{pages}{072502} (\bibinfo{year}{2008}).

\bibitem[{\citenamefont{Mart{\'\i}n-Sol{\'\i}s
  et~al.}(2017)\citenamefont{Mart{\'\i}n-Sol{\'\i}s, Loarte, and
  Lehnen}}]{Martin:2017}
\bibinfo{author}{\bibfnamefont{J.}~\bibnamefont{Mart{\'\i}n-Sol{\'\i}s}},
  \bibinfo{author}{\bibfnamefont{A.}~\bibnamefont{Loarte}}, \bibnamefont{and}
  \bibinfo{author}{\bibfnamefont{M.}~\bibnamefont{Lehnen}},
  \bibinfo{journal}{Nuclear Fusion} \textbf{\bibinfo{volume}{57}},
  \bibinfo{pages}{066025} (\bibinfo{year}{2017}).

\bibitem[{\citenamefont{Aleynikov and Breizman}(2015)}]{Aleynikov:2015}
\bibinfo{author}{\bibfnamefont{P.}~\bibnamefont{Aleynikov}} \bibnamefont{and}
  \bibinfo{author}{\bibfnamefont{B.~N.} \bibnamefont{Breizman}},
  \bibinfo{journal}{Physical review letters} \textbf{\bibinfo{volume}{114}},
  \bibinfo{pages}{155001} (\bibinfo{year}{2015}).

\bibitem[{\citenamefont{Hesslow
  et~al.}(2019{\natexlab{a}})\citenamefont{Hesslow, Embr{\'e}us, Vallhagen, and
  F{\"u}l{\"o}p}}]{hesslow2019influence}
\bibinfo{author}{\bibfnamefont{L.}~\bibnamefont{Hesslow}},
  \bibinfo{author}{\bibfnamefont{O.}~\bibnamefont{Embr{\'e}us}},
  \bibinfo{author}{\bibfnamefont{O.}~\bibnamefont{Vallhagen}},
  \bibnamefont{and}
  \bibinfo{author}{\bibfnamefont{T.}~\bibnamefont{F{\"u}l{\"o}p}},
  \bibinfo{journal}{Nuclear Fusion} \textbf{\bibinfo{volume}{59}},
  \bibinfo{pages}{084004} (\bibinfo{year}{2019}{\natexlab{a}}).

\bibitem[{\citenamefont{McDevitt et~al.}(2018)\citenamefont{McDevitt, Guo, and
  Tang}}]{mcdevitt2018relation}
\bibinfo{author}{\bibfnamefont{C.~J.} \bibnamefont{McDevitt}},
  \bibinfo{author}{\bibfnamefont{Z.}~\bibnamefont{Guo}}, \bibnamefont{and}
  \bibinfo{author}{\bibfnamefont{X.-Z.} \bibnamefont{Tang}},
  \bibinfo{journal}{Plasma Physics and Controlled Fusion}
  \textbf{\bibinfo{volume}{60}}, \bibinfo{pages}{024004}
  (\bibinfo{year}{2018}).

\bibitem[{\citenamefont{Hesslow
  et~al.}(2019{\natexlab{b}})\citenamefont{Hesslow, Unnerfelt, Vallhagen,
  Embr{\'e}us, Hoppe, Papp, and F{\"u}l{\"o}p}}]{hesslow2019evaluation}
\bibinfo{author}{\bibfnamefont{L.}~\bibnamefont{Hesslow}},
  \bibinfo{author}{\bibfnamefont{L.}~\bibnamefont{Unnerfelt}},
  \bibinfo{author}{\bibfnamefont{O.}~\bibnamefont{Vallhagen}},
  \bibinfo{author}{\bibfnamefont{O.}~\bibnamefont{Embr{\'e}us}},
  \bibinfo{author}{\bibfnamefont{M.}~\bibnamefont{Hoppe}},
  \bibinfo{author}{\bibfnamefont{G.}~\bibnamefont{Papp}}, \bibnamefont{and}
  \bibinfo{author}{\bibfnamefont{T.}~\bibnamefont{F{\"u}l{\"o}p}},
  \bibinfo{journal}{Journal of Plasma Physics} \textbf{\bibinfo{volume}{85}}
  (\bibinfo{year}{2019}{\natexlab{b}}).

\bibitem[{\citenamefont{McDevitt}(2023)}]{McDevitt:hottail:2023}
\bibinfo{author}{\bibfnamefont{C.~J.} \bibnamefont{McDevitt}},
  \bibinfo{journal}{Physics of Plasmas} \textbf{\bibinfo{volume}{30}},
  \bibinfo{pages}{092501} (\bibinfo{year}{2023}), ISSN
  \bibinfo{issn}{1070-664X}, \urlprefix\url{https://doi.org/10.1063/5.0164712}.

\bibitem[{\citenamefont{Yang et~al.}(2024)\citenamefont{Yang, Wang, del
  Castillo-Negrete, Cao, and Zhang}}]{yang2024pseudoreversible}
\bibinfo{author}{\bibfnamefont{M.}~\bibnamefont{Yang}},
  \bibinfo{author}{\bibfnamefont{P.}~\bibnamefont{Wang}},
  \bibinfo{author}{\bibfnamefont{D.}~\bibnamefont{del Castillo-Negrete}},
  \bibinfo{author}{\bibfnamefont{Y.}~\bibnamefont{Cao}}, \bibnamefont{and}
  \bibinfo{author}{\bibfnamefont{G.}~\bibnamefont{Zhang}},
  \bibinfo{journal}{SIAM Journal on Scientific Computing}
  \textbf{\bibinfo{volume}{46}}, \bibinfo{pages}{C508} (\bibinfo{year}{2024}).

\bibitem[{\citenamefont{Arnaud et~al.}(2024)\citenamefont{Arnaud, Mark, and
  McDevitt}}]{arnaud2024physics}
\bibinfo{author}{\bibfnamefont{J.~S.} \bibnamefont{Arnaud}},
  \bibinfo{author}{\bibfnamefont{T.}~\bibnamefont{Mark}}, \bibnamefont{and}
  \bibinfo{author}{\bibfnamefont{C.~J.} \bibnamefont{McDevitt}},
  \bibinfo{journal}{J. Plasma Phys.} \textbf{\bibinfo{volume}{90}},
  \bibinfo{pages}{905900409} (\bibinfo{year}{2024}).

\bibitem[{\citenamefont{Harvey et~al.}(2000)\citenamefont{Harvey, Chan, Chiu,
  Evans, Rosenbluth, and Whyte}}]{Harvey:2000}
\bibinfo{author}{\bibfnamefont{R.}~\bibnamefont{Harvey}},
  \bibinfo{author}{\bibfnamefont{V.}~\bibnamefont{Chan}},
  \bibinfo{author}{\bibfnamefont{S.}~\bibnamefont{Chiu}},
  \bibinfo{author}{\bibfnamefont{T.}~\bibnamefont{Evans}},
  \bibinfo{author}{\bibfnamefont{M.}~\bibnamefont{Rosenbluth}},
  \bibnamefont{and} \bibinfo{author}{\bibfnamefont{D.}~\bibnamefont{Whyte}},
  \bibinfo{journal}{Physics of Plasmas} \textbf{\bibinfo{volume}{7}},
  \bibinfo{pages}{4590} (\bibinfo{year}{2000}).

\bibitem[{\citenamefont{Nilsson et~al.}(2015)\citenamefont{Nilsson, Decker,
  Peysson, Granetz, Saint-Laurent, and Vlainic}}]{Nilsson:2015}
\bibinfo{author}{\bibfnamefont{E.}~\bibnamefont{Nilsson}},
  \bibinfo{author}{\bibfnamefont{J.}~\bibnamefont{Decker}},
  \bibinfo{author}{\bibfnamefont{Y.}~\bibnamefont{Peysson}},
  \bibinfo{author}{\bibfnamefont{R.~S.} \bibnamefont{Granetz}},
  \bibinfo{author}{\bibfnamefont{F.}~\bibnamefont{Saint-Laurent}},
  \bibnamefont{and} \bibinfo{author}{\bibfnamefont{M.}~\bibnamefont{Vlainic}},
  \bibinfo{journal}{Plasma Physics and Controlled Fusion}
  \textbf{\bibinfo{volume}{57}}, \bibinfo{pages}{095006}
  (\bibinfo{year}{2015}).

\bibitem[{\citenamefont{Guo et~al.}(2019)\citenamefont{Guo, Mcdevitt, and
  Tang}}]{guo-etal-pop-2019}
\bibinfo{author}{\bibfnamefont{Z.}~\bibnamefont{Guo}},
  \bibinfo{author}{\bibfnamefont{C.}~\bibnamefont{Mcdevitt}}, \bibnamefont{and}
  \bibinfo{author}{\bibfnamefont{X.}~\bibnamefont{Tang}},
  \bibinfo{journal}{Physics of Plasmas} \textbf{\bibinfo{volume}{26}},
  \bibinfo{pages}{082503} (\bibinfo{year}{2019}), ISSN
  \bibinfo{issn}{1070-664X}, \bibinfo{note}{\_eprint:
  https://pubs.aip.org/aip/pop/article-pdf/doi/10.1063/1.5055874/15853638/082503\_1\_online.pdf},
  \urlprefix\url{https://doi.org/10.1063/1.5055874}.

\bibitem[{\citenamefont{Hoppe et~al.}(2021)\citenamefont{Hoppe, Embreus, and
  F{\"u}l{\"o}p}}]{hoppe2021dream}
\bibinfo{author}{\bibfnamefont{M.}~\bibnamefont{Hoppe}},
  \bibinfo{author}{\bibfnamefont{O.}~\bibnamefont{Embreus}}, \bibnamefont{and}
  \bibinfo{author}{\bibfnamefont{T.}~\bibnamefont{F{\"u}l{\"o}p}},
  \bibinfo{journal}{Computer Physics Communications}
  \textbf{\bibinfo{volume}{268}}, \bibinfo{pages}{108098}
  (\bibinfo{year}{2021}).

\bibitem[{\citenamefont{Rudi et~al.}(2024)\citenamefont{Rudi, Heldman,
  Constantinescu, Tang, and Tang}}]{Rudi-etal-JCP-2024}
\bibinfo{author}{\bibfnamefont{J.}~\bibnamefont{Rudi}},
  \bibinfo{author}{\bibfnamefont{M.}~\bibnamefont{Heldman}},
  \bibinfo{author}{\bibfnamefont{E.~M.} \bibnamefont{Constantinescu}},
  \bibinfo{author}{\bibfnamefont{Q.}~\bibnamefont{Tang}}, \bibnamefont{and}
  \bibinfo{author}{\bibfnamefont{X.-Z.} \bibnamefont{Tang}},
  \bibinfo{journal}{Journal of Computational Physics}
  \textbf{\bibinfo{volume}{507}}, \bibinfo{pages}{112954}
  (\bibinfo{year}{2024}), ISSN \bibinfo{issn}{0021-9991},
  \urlprefix\url{https://www.sciencedirect.com/science/article/pii/S0021999124002031}.

\bibitem[{\citenamefont{Eriksson and Helander}(2003)}]{Eriksson:2003}
\bibinfo{author}{\bibfnamefont{L.-G.} \bibnamefont{Eriksson}} \bibnamefont{and}
  \bibinfo{author}{\bibfnamefont{P.}~\bibnamefont{Helander}},
  \bibinfo{journal}{Computer Physics Communications}
  \textbf{\bibinfo{volume}{154}}, \bibinfo{pages}{175} (\bibinfo{year}{2003}).

\bibitem[{\citenamefont{Sommariva et~al.}(2017)\citenamefont{Sommariva, Nardon,
  Beyer, Hoelzl, Huijsmans, van Vugt, and Contributors}}]{Sommariva:2017}
\bibinfo{author}{\bibfnamefont{C.}~\bibnamefont{Sommariva}},
  \bibinfo{author}{\bibfnamefont{E.}~\bibnamefont{Nardon}},
  \bibinfo{author}{\bibfnamefont{P.}~\bibnamefont{Beyer}},
  \bibinfo{author}{\bibfnamefont{M.}~\bibnamefont{Hoelzl}},
  \bibinfo{author}{\bibfnamefont{G.}~\bibnamefont{Huijsmans}},
  \bibinfo{author}{\bibfnamefont{D.}~\bibnamefont{van Vugt}}, \bibnamefont{and}
  \bibinfo{author}{\bibfnamefont{J.}~\bibnamefont{Contributors}},
  \bibinfo{journal}{Nuclear Fusion} \textbf{\bibinfo{volume}{58}},
  \bibinfo{pages}{016043} (\bibinfo{year}{2017}).

\bibitem[{\citenamefont{McDevitt
  et~al.}(2019{\natexlab{a}})\citenamefont{McDevitt, Guo, and
  Tang}}]{mcdevitt2019avalanche}
\bibinfo{author}{\bibfnamefont{C.~J.} \bibnamefont{McDevitt}},
  \bibinfo{author}{\bibfnamefont{Z.}~\bibnamefont{Guo}}, \bibnamefont{and}
  \bibinfo{author}{\bibfnamefont{X.~Z.} \bibnamefont{Tang}},
  \bibinfo{journal}{Plasma Physics and Controlled Fusion}
  \textbf{\bibinfo{volume}{61}}, \bibinfo{pages}{054008}
  (\bibinfo{year}{2019}{\natexlab{a}}).

\bibitem[{\citenamefont{Beidler et~al.}(2024)\citenamefont{Beidler, del
  Castillo-Negrete, Shiraki, Baylor, Hollmann, and Lasnier}}]{beidler2024wall}
\bibinfo{author}{\bibfnamefont{M.}~\bibnamefont{Beidler}},
  \bibinfo{author}{\bibfnamefont{D.}~\bibnamefont{del Castillo-Negrete}},
  \bibinfo{author}{\bibfnamefont{D.}~\bibnamefont{Shiraki}},
  \bibinfo{author}{\bibfnamefont{L.~R.} \bibnamefont{Baylor}},
  \bibinfo{author}{\bibfnamefont{E.~M.} \bibnamefont{Hollmann}},
  \bibnamefont{and} \bibinfo{author}{\bibfnamefont{C.}~\bibnamefont{Lasnier}},
  \bibinfo{journal}{Nuclear Fusion} \textbf{\bibinfo{volume}{64}},
  \bibinfo{pages}{076038} (\bibinfo{year}{2024}).

\bibitem[{\citenamefont{SOKOLOV}(1979)}]{Sokolov:1979}
\bibinfo{author}{\bibfnamefont{I.}~\bibnamefont{SOKOLOV}},
  \bibinfo{journal}{JETP Letters} \textbf{\bibinfo{volume}{29}},
  \bibinfo{pages}{218} (\bibinfo{year}{1979}).

\bibitem[{\citenamefont{Hender et~al.}(2007)\citenamefont{Hender, Wesley,
  Bialek, Bondeson, Boozer, Buttery, Garofalo, Goodman, Granetz, Gribov
  et~al.}}]{Hender:2007}
\bibinfo{author}{\bibfnamefont{T.}~\bibnamefont{Hender}},
  \bibinfo{author}{\bibfnamefont{J.}~\bibnamefont{Wesley}},
  \bibinfo{author}{\bibfnamefont{J.}~\bibnamefont{Bialek}},
  \bibinfo{author}{\bibfnamefont{A.}~\bibnamefont{Bondeson}},
  \bibinfo{author}{\bibfnamefont{A.}~\bibnamefont{Boozer}},
  \bibinfo{author}{\bibfnamefont{R.}~\bibnamefont{Buttery}},
  \bibinfo{author}{\bibfnamefont{A.}~\bibnamefont{Garofalo}},
  \bibinfo{author}{\bibfnamefont{T.}~\bibnamefont{Goodman}},
  \bibinfo{author}{\bibfnamefont{R.}~\bibnamefont{Granetz}},
  \bibinfo{author}{\bibfnamefont{Y.}~\bibnamefont{Gribov}},
  \bibnamefont{et~al.}, \bibinfo{journal}{Nuclear fusion}
  \textbf{\bibinfo{volume}{47}}, \bibinfo{pages}{S128} (\bibinfo{year}{2007}).

\bibitem[{\citenamefont{Breizman et~al.}(2019)\citenamefont{Breizman,
  Aleynikov, Hollmann, and Lehnen}}]{breizman2019physics}
\bibinfo{author}{\bibfnamefont{B.~N.} \bibnamefont{Breizman}},
  \bibinfo{author}{\bibfnamefont{P.}~\bibnamefont{Aleynikov}},
  \bibinfo{author}{\bibfnamefont{E.~M.} \bibnamefont{Hollmann}},
  \bibnamefont{and} \bibinfo{author}{\bibfnamefont{M.}~\bibnamefont{Lehnen}},
  \bibinfo{journal}{Nuclear Fusion} \textbf{\bibinfo{volume}{59}},
  \bibinfo{pages}{083001} (\bibinfo{year}{2019}).

\bibitem[{\citenamefont{Liu et~al.}(2017)\citenamefont{Liu, Brennan, Boozer,
  and Bhattacharjee}}]{Liu:2017}
\bibinfo{author}{\bibfnamefont{C.}~\bibnamefont{Liu}},
  \bibinfo{author}{\bibfnamefont{D.~P.} \bibnamefont{Brennan}},
  \bibinfo{author}{\bibfnamefont{A.~H.} \bibnamefont{Boozer}},
  \bibnamefont{and}
  \bibinfo{author}{\bibfnamefont{A.}~\bibnamefont{Bhattacharjee}},
  \bibinfo{journal}{Plasma Physics and Controlled Fusion}
  \textbf{\bibinfo{volume}{59}}, \bibinfo{pages}{024003}
  (\bibinfo{year}{2017}),
  \urlprefix\url{http://stacks.iop.org/0741-3335/59/i=2/a=024003}.

\bibitem[{\citenamefont{Hesslow et~al.}(2017)\citenamefont{Hesslow, Embr\'eus,
  Stahl, DuBois, Papp, Newton, and F\"ul\"op}}]{Hesslow:2017}
\bibinfo{author}{\bibfnamefont{L.}~\bibnamefont{Hesslow}},
  \bibinfo{author}{\bibfnamefont{O.}~\bibnamefont{Embr\'eus}},
  \bibinfo{author}{\bibfnamefont{A.}~\bibnamefont{Stahl}},
  \bibinfo{author}{\bibfnamefont{T.~C.} \bibnamefont{DuBois}},
  \bibinfo{author}{\bibfnamefont{G.}~\bibnamefont{Papp}},
  \bibinfo{author}{\bibfnamefont{S.~L.} \bibnamefont{Newton}},
  \bibnamefont{and}
  \bibinfo{author}{\bibfnamefont{T.}~\bibnamefont{F\"ul\"op}},
  \bibinfo{journal}{Phys. Rev. Lett.} \textbf{\bibinfo{volume}{118}},
  \bibinfo{pages}{255001} (\bibinfo{year}{2017}),
  \urlprefix\url{https://link.aps.org/doi/10.1103/PhysRevLett.118.255001}.

\bibitem[{\citenamefont{Karney and Fisch}(1986)}]{karney1986current}
\bibinfo{author}{\bibfnamefont{C.~F.} \bibnamefont{Karney}} \bibnamefont{and}
  \bibinfo{author}{\bibfnamefont{N.~J.} \bibnamefont{Fisch}},
  \bibinfo{journal}{The Physics of fluids} \textbf{\bibinfo{volume}{29}},
  \bibinfo{pages}{180} (\bibinfo{year}{1986}).

\bibitem[{\citenamefont{Zhang and del
  Castillo-Negrete}(2017)}]{zhang2017backward}
\bibinfo{author}{\bibfnamefont{G.}~\bibnamefont{Zhang}} \bibnamefont{and}
  \bibinfo{author}{\bibfnamefont{D.}~\bibnamefont{del Castillo-Negrete}},
  \bibinfo{journal}{Physics of Plasmas} \textbf{\bibinfo{volume}{24}},
  \bibinfo{pages}{092511} (\bibinfo{year}{2017}).

\bibitem[{\citenamefont{M{\o}ller}(1932)}]{Moller:1932}
\bibinfo{author}{\bibfnamefont{C.}~\bibnamefont{M{\o}ller}},
  \bibinfo{journal}{Annalen der Physik} \textbf{\bibinfo{volume}{406}},
  \bibinfo{pages}{531} (\bibinfo{year}{1932}).

\bibitem[{\citenamefont{Ashkin et~al.}(1954)\citenamefont{Ashkin, Page, and
  Woodward}}]{Ashkin:1954}
\bibinfo{author}{\bibfnamefont{A.}~\bibnamefont{Ashkin}},
  \bibinfo{author}{\bibfnamefont{L.~A.} \bibnamefont{Page}}, \bibnamefont{and}
  \bibinfo{author}{\bibfnamefont{W.}~\bibnamefont{Woodward}},
  \bibinfo{journal}{Physical Review} \textbf{\bibinfo{volume}{94}},
  \bibinfo{pages}{357} (\bibinfo{year}{1954}).

\bibitem[{\citenamefont{Boozer}(2015)}]{Boozer:2015}
\bibinfo{author}{\bibfnamefont{A.~H.} \bibnamefont{Boozer}},
  \bibinfo{journal}{Physics of Plasmas} \textbf{\bibinfo{volume}{22}},
  \bibinfo{pages}{032504} (\bibinfo{year}{2015}).

\bibitem[{\citenamefont{Raissi et~al.}(2019)\citenamefont{Raissi, Perdikaris,
  and Karniadakis}}]{raissi2019physics}
\bibinfo{author}{\bibfnamefont{M.}~\bibnamefont{Raissi}},
  \bibinfo{author}{\bibfnamefont{P.}~\bibnamefont{Perdikaris}},
  \bibnamefont{and} \bibinfo{author}{\bibfnamefont{G.~E.}
  \bibnamefont{Karniadakis}}, \bibinfo{journal}{Journal of Computational
  physics} \textbf{\bibinfo{volume}{378}}, \bibinfo{pages}{686}
  (\bibinfo{year}{2019}).

\bibitem[{\citenamefont{Karniadakis et~al.}(2021)\citenamefont{Karniadakis,
  Kevrekidis, Lu, Perdikaris, Wang, and Yang}}]{karniadakis2021physics}
\bibinfo{author}{\bibfnamefont{G.~E.} \bibnamefont{Karniadakis}},
  \bibinfo{author}{\bibfnamefont{I.~G.} \bibnamefont{Kevrekidis}},
  \bibinfo{author}{\bibfnamefont{L.}~\bibnamefont{Lu}},
  \bibinfo{author}{\bibfnamefont{P.}~\bibnamefont{Perdikaris}},
  \bibinfo{author}{\bibfnamefont{S.}~\bibnamefont{Wang}}, \bibnamefont{and}
  \bibinfo{author}{\bibfnamefont{L.}~\bibnamefont{Yang}},
  \bibinfo{journal}{Nature Reviews Physics} \textbf{\bibinfo{volume}{3}},
  \bibinfo{pages}{422} (\bibinfo{year}{2021}).

\bibitem[{\citenamefont{Lu et~al.}(2021)\citenamefont{Lu, Meng, Mao, and
  Karniadakis}}]{lu2021deepxde}
\bibinfo{author}{\bibfnamefont{L.}~\bibnamefont{Lu}},
  \bibinfo{author}{\bibfnamefont{X.}~\bibnamefont{Meng}},
  \bibinfo{author}{\bibfnamefont{Z.}~\bibnamefont{Mao}}, \bibnamefont{and}
  \bibinfo{author}{\bibfnamefont{G.~E.} \bibnamefont{Karniadakis}},
  \bibinfo{journal}{SIAM Review} \textbf{\bibinfo{volume}{63}},
  \bibinfo{pages}{208} (\bibinfo{year}{2021}).

\bibitem[{\citenamefont{Abadi et~al.}(2016)\citenamefont{Abadi, Barham, Chen,
  Chen, Davis, Dean, Devin, Ghemawat, Irving, Isard
  et~al.}}]{abadi2016tensorflow}
\bibinfo{author}{\bibfnamefont{M.}~\bibnamefont{Abadi}},
  \bibinfo{author}{\bibfnamefont{P.}~\bibnamefont{Barham}},
  \bibinfo{author}{\bibfnamefont{J.}~\bibnamefont{Chen}},
  \bibinfo{author}{\bibfnamefont{Z.}~\bibnamefont{Chen}},
  \bibinfo{author}{\bibfnamefont{A.}~\bibnamefont{Davis}},
  \bibinfo{author}{\bibfnamefont{J.}~\bibnamefont{Dean}},
  \bibinfo{author}{\bibfnamefont{M.}~\bibnamefont{Devin}},
  \bibinfo{author}{\bibfnamefont{S.}~\bibnamefont{Ghemawat}},
  \bibinfo{author}{\bibfnamefont{G.}~\bibnamefont{Irving}},
  \bibinfo{author}{\bibfnamefont{M.}~\bibnamefont{Isard}},
  \bibnamefont{et~al.}, in \emph{\bibinfo{booktitle}{12th USENIX symposium on
  operating systems design and implementation (OSDI 16)}}
  (\bibinfo{year}{2016}), pp. \bibinfo{pages}{265--283}.

\bibitem[{\citenamefont{Wu et~al.}(2023)\citenamefont{Wu, Zhu, Tan, Kartha, and
  Lu}}]{wu2023comprehensive}
\bibinfo{author}{\bibfnamefont{C.}~\bibnamefont{Wu}},
  \bibinfo{author}{\bibfnamefont{M.}~\bibnamefont{Zhu}},
  \bibinfo{author}{\bibfnamefont{Q.}~\bibnamefont{Tan}},
  \bibinfo{author}{\bibfnamefont{Y.}~\bibnamefont{Kartha}}, \bibnamefont{and}
  \bibinfo{author}{\bibfnamefont{L.}~\bibnamefont{Lu}},
  \bibinfo{journal}{Computer Methods in Applied Mechanics and Engineering}
  \textbf{\bibinfo{volume}{403}}, \bibinfo{pages}{115671}
  (\bibinfo{year}{2023}).

\bibitem[{\citenamefont{Chiu et~al.}(1998)\citenamefont{Chiu, Rosenbluth,
  Harvey, and Chan}}]{Chiu:1998}
\bibinfo{author}{\bibfnamefont{S.}~\bibnamefont{Chiu}},
  \bibinfo{author}{\bibfnamefont{M.}~\bibnamefont{Rosenbluth}},
  \bibinfo{author}{\bibfnamefont{R.}~\bibnamefont{Harvey}}, \bibnamefont{and}
  \bibinfo{author}{\bibfnamefont{V.}~\bibnamefont{Chan}},
  \bibinfo{journal}{Nuclear Fusion} \textbf{\bibinfo{volume}{38}},
  \bibinfo{pages}{1711} (\bibinfo{year}{1998}).

\bibitem[{\citenamefont{McDevitt and Tang}(2019)}]{mcdevitt2019runaway}
\bibinfo{author}{\bibfnamefont{C.}~\bibnamefont{McDevitt}} \bibnamefont{and}
  \bibinfo{author}{\bibfnamefont{X.-Z.} \bibnamefont{Tang}},
  \bibinfo{journal}{EPL (Europhysics Letters)} \textbf{\bibinfo{volume}{127}},
  \bibinfo{pages}{45001} (\bibinfo{year}{2019}).

\bibitem[{\citenamefont{Arnaud and McDevitt}(2024)}]{arnaud2024impact}
\bibinfo{author}{\bibfnamefont{J.~S.} \bibnamefont{Arnaud}} \bibnamefont{and}
  \bibinfo{author}{\bibfnamefont{C.~J.} \bibnamefont{McDevitt}},
  \bibinfo{journal}{Physics of Plasmas} \textbf{\bibinfo{volume}{31}}
  (\bibinfo{year}{2024}).

\bibitem[{\citenamefont{McDevitt
  et~al.}(2019{\natexlab{b}})\citenamefont{McDevitt, Guo, and
  Tang}}]{mcdevitt2019spatial}
\bibinfo{author}{\bibfnamefont{C.~J.} \bibnamefont{McDevitt}},
  \bibinfo{author}{\bibfnamefont{Z.}~\bibnamefont{Guo}}, \bibnamefont{and}
  \bibinfo{author}{\bibfnamefont{X.-Z.} \bibnamefont{Tang}},
  \bibinfo{journal}{Plasma Physics and Controlled Fusion}
  \textbf{\bibinfo{volume}{61}}, \bibinfo{pages}{024004}
  (\bibinfo{year}{2019}{\natexlab{b}}).

\end{thebibliography}

\end{document}